%% file: main.tex
\documentclass[conference]{IEEEtran}
\IEEEoverridecommandlockouts
\usepackage{cite}
\usepackage{amsmath,amssymb,amsfonts}
\usepackage{algorithm}
\usepackage{algorithmic}

\usepackage{graphicx}
\usepackage{textcomp}
\usepackage{xcolor}
\usepackage{balance,color}
\usepackage{multirow}
\usepackage{amsmath}
\usepackage{subcaption}
\usepackage{tikz}
\usepackage{hyperref}
\newcommand{\ignore}[1]{}

\usepackage{fancyhdr}
\newcommand{\linebreakand}{%
  \begin{@IEEEauthorhalign}
  \hfill\mbox{}\par
  \mbox{}\hfill\end{@IEEEauthorhalign}
}

\def\BibTeX{{\rm B\kern-.05em{\sc i\kern-.025em b}\kern-.08em
    T\kern-.1667em\lower.7ex\hbox{E}\kern-.125emX}}

\makeatother
\begin{document}
\pagestyle{plain}
%\pagestyle{fancy}

%\title{Visually Analyze SHAP Explanation Plots to Identify False Positives and False Negatives in Intrusion Detection}

\title{Visually Analyze SHAP Plots to Diagnose Misclassifications in ML-based Intrusion Detection}

\author{
    \IEEEauthorblockN{Maraz Mia}
    \IEEEauthorblockA{
        \textit{Department of Computer Science, Tennessee Tech} \\
        Cookeville, TN, USA \\
        mmia43@tntech.edu
    }
    
    \linebreakand

    \IEEEauthorblockN{Tariqul Islam}
    \IEEEauthorblockA{
        \textit{School of Information Studies, Syracuse University} \\
        Syracuse, NY, USA \\
        mtislam@syr.edu
    }

    % \vspace{-4mm}
    \and

    \IEEEauthorblockN{Mir Mehedi A. Pritom}
    \IEEEauthorblockA{
        \textit{Department of Computer Science, Tennessee Tech} \\
        Cookeville, TN, USA \\
        mpritom@tntech.edu
    }
    
    % \and
     \vspace{4.5mm}
    
    \IEEEauthorblockN{Kamrul Hasan}
    \IEEEauthorblockA{
        \textit{Department of ECE, Tennessee State University} \\
        Nashville, TN, USA \\
        mhasan1@tnstate.edu
    }
}

\maketitle

\thispagestyle{fancy}
\fancyhead[L]{This paper is accepted at the MLC Workshop in the International Conference on Data Mining (ICDM 2024)}

\fancyhead[L]{This paper is accepted at the MLC Workshop in the International Conference on Data Mining Conference (ICDM 2024)}

\begin{abstract}
Intrusion detection has been a commonly adopted detective security measures to safeguard systems and networks from various threats. A robust intrusion detection system (IDS) can essentially mitigate threats by providing alerts. In networks based IDS, typically we deal with cyber threats like distributed denial of service (DDoS), spoofing, reconnaissance, brute-force, botnets, and so on. In order to detect these threats various machine learning (ML) and deep learning (DL) models have been proposed. However, one of the key challenges with these predictive approaches is the presence of false positive (FP) and false negative (FN) instances. This FPs and FNs within any black-box intrusion detection system (IDS) make the decision-making task of an analyst further complicated. In this paper, we propose an explainable artificial intelligence (XAI) based visual analysis approach using overlapping SHAP plots that presents the feature explanation to identify potential false positive and false negatives in IDS. Our approach can further provide guidance to security analysts for effective decision-making. We present case study with multiple publicly available network traffic datasets to showcase the efficacy of our approach for identifying false positive and false negative instances. Our use-case scenarios provide clear guidance for analysts on how to use the visual analysis approach for reliable course-of-actions against such threats. %case study results show that our approach is very competitive with existing literature and would certainly help analyst significantly to identify false-positives and false-negatives instances detected by any IDS.
\end{abstract}


\vspace{-1mm}
\section{Introduction}
\label{sec:intro}
\input{introduction.tex}

\vspace{-2mm}
\section{Related Works}

%\footnote{review comment 2: - related work section lacks analysis of existing studies, some citations are too long.}
\label{sec:related-work}
\input{related-work.tex}

\vspace{-1.5mm}
\section{Methodology}
\label{sec:method}
\input{methodology.tex}

\vspace{-1.5mm}
\section{Case Study Results and Use Case Scenarios}
\label{sec:case_study}
\input{experiment-results}

\vspace{-2mm}
\section{Discussion}
\label{sec:discuss}
\input{discuss}% \vspace{-1em}

\vspace{-2mm}
\section{Conclusion}
\label{sec:conclusion}
\input{conclusion.tex}% \vspace{-1em}

%\section*{Acknowledgment}

\vspace{-3mm}
%\section*{References}
\bibliographystyle{IEEEtran}
\bibliography{mybibliography}

\newpage
\section*{Appendix}
\label{sec:appendix}
\input{appendix.tex}% \vspace{-1em}

\end{document}

%% file: introduction.tex
\vspace{-0.5em}
Intrusion detection systems are ubiquitous in network and commuter systems which check on every request and response over a computer or network and examines for indications of potential cyber attacks or threats, including attempts for exploitation and other situations that poses an immediate threat to the network \cite{IDS1}. Enhancing the effectiveness of current IDS is challenging and crucial for detective and preventive cyber defense. As the digital era progresses, computer systems and networks including internet of things (IoT) are vulnerable to more sophisticated attacks. %The Internet of Things (IoT) is the infrastructure that will make it possible for collaborative objects and conventional networks to function together seamlessly. 
Alike the traditional sensor networks, IoT network also has data traffic to be shared among multiple IoT devices such as smart home (e.g., Google Home, Amazon Echo), health monitoring (e.g., DexCom glucose monitor), wearable (e.g., Fitbit), smart manufacturing (e.g., collaborative robots), smart agriculture (e.g., soil sensors) or smart retail (e.g., Beacons, smart shelves) those are vulnerable to various cyber threats. %The quantity of data that travels across a network at any given time is referred to as network traffic, or data traffic, in Internet of Things devices. are typically sent as data packets that are received over a network and constructed again on the receiving end. 
%Certain types of devices with sensors are used to gather data from tangible objects. 
%With the current technological advancements, attackers are propagating various types of complex and zero-day attacks within the IoT network, which can be challenging to detect without proper predictive analytics. %In addition, power consumption, scalability, and stability are a few additional important IIoT (Industrial Internet of Things) concerns. %In certain cases, conventional security methods are not suitable.
%According to {\color{red}?? report \cite{}}, 20\% of the affected enterprises fall under the compact business class, 33\% fall under the small business class, and 41\% fall under the large corporation class which indicates that more effective IDSes are required for active cyber defense. A minimum of three instances of data theft have left the services provided by 82\% of organizations' clients worthless. A 26\% decrease in operational efficiency and a 41\% disruption of the targeted services were reported by the organizations affected by DDoS assaults \cite{Behal2017CharacterizationAC}.
The most common types of attacks observed within the IoT environments are Distributed Denial of Service (DDoS), Denial of Service (DoS), brute force attacks, spoofing attacks, website-based attacks (e.g., XSS, SQL injection, defacement), man-in-the-middle attacks, replay attacks, network reconnaissance, and Mirai botnets \cite{IDS-attacks}. To identify such attacks in any network, researchers have long since worked to come up with detection mechanisms that can be both adaptive to new types of attacks and also be more practical in real-world scenarios. The more promising motivations behind the usage of machine learning or other rule-based intrusion detection systems is that it reduces operational overhead and human-centric errors \cite{Ontologybased}. However, this also creates a backdoor for misclassification-- either false positives or false negatives. Nonetheless, most of the state-of-the-art network intrusion detection systems (NIDS), host-based intrusion detection systems (HIDS), or log analysis \cite{mir_host_loganalysis_2017} rely on black-box machine learning model-based prediction, which suffer from false positives and false negatives identifications. %and thus questions the effectiveness of the system.  %iden  model-based detection of IoT intrusions that can also help reduce the {\em false positives} and {\em false negatives} by distinguishing the feature-based explanations for true positive, false positive, true negative, and false negative groups. 

As most of the existing methods require decision from an analyst \cite{XAI_IDS_7, XAI_IDS_8, pritom2022supporting} to culminate the prediction of an ML model, it is expected that the analyst must be aided with corresponding features' contribution to make trustworthy decisions. Depending on the knowledge capability of the analyst, a visual characteristics of the misclassification cases (FPs and FNs) can be an effective approach, which is not systematically addressed in the existing literature for decision making. % in intrusion detection scenarios. %However, this method (simple but effective and less time and domain knowledge dependent) has not yet been recommended by the research community. }{
%There are some existing work on FP and FN reduction for ML based detection .....[discuss the motivation of our research here by citing existing work that can be compared to our work, where do we stand? why our research is important and novel?]
%\footnote{We will address here the Reviewer's comment 1: there are many similar studies on the topic, but not all of them were analyzed. The motivation needs more analyses.}
In this paper, we propose an explainable AI based intrusion detection and present a new step-by-step methodology for using SHAP feature explanation plots \cite{shap} by the analysts for potentially identifying false positives and false negatives. We present our methodological approach with empirical case studies on multiple publicly available network traffic datasets. %published by the Canadian Institute of Cybersecurity (CIC) including IoT traffic dataset \cite{CICIoT-2023-datasetpaper} and from other sources. %we have used a recent IoT dataset published by the Canadian Institute of Cybersecurity (CIC-IoT-2023)  and empirically show that our methods have achieved improved overall performance for the local explanation (e.g, using SHAP \cite{shap}) based detection of intrusions as well as having more useful information for reducing false positives and false negatives to aid the decision-making process. %for any alerts generated for the detection.  %comparatively and analytically on the existing and the most updated ML-based Intrusion Detection Systems (NIDS) that used CIC IoT 2023 dataset \cite{s23135941} as their research background, and also provide some analogies to quantify their approach based on some empirical data analysis and basic feature engineering and ML concepts. 
%Additionally, the CIC-IoT-2023 dataset has a class imbalance problem, which is pretty common in IDS datasets to impact the results with training bias. So, we address the data imbalance by running empirical case studies for both SMOTE (Synthetic Minority Oversampling Technique) \cite{SMOTE_Chawla_2002} and random under-sampling methods \cite{rand_under_sample} to pick the best approach and create an IDS model that provides more generalized results in a real-world dynamic environment. 
We also discuss the usage of {\em Brier score} \cite{rufibach2010use_brier} as a reliable metric of confidence for various ML  model's performance evaluation when tested for classification of attack versus benign traffic within the datasets. {\em Brier Score} allows us to interpret how close the model predicted raw probability values to the actual outcomes. However, the black-box nature of models can not be addressed by {\em Brier Score}, and thus feature based SHAP explanation is necessary to interpret why a particular label is predicted for an individual traffic data-point. In summary, we have made the following major contributions in this paper: %Finally, we introduced XAI enabled FP and FN mitigation techniques and provides valuable insights of the feature contribution in those cases using SHAP (SHapley Additive exPlanations).

% {\color{red}\footnote{this should go as bullet point list in the Intro...\color{green}{Done}}
%Although, in all previous studies, researchers showed their results in terms of accuracy or other metrics, there are some crucial information and evaluations still missing. Some important concerns are :
% \footnote{re-write the contributions list based on the above arguments....I think it should focus on providing local explnation for individual identification, you can mention the false positive and false negative group explanations as well to identify them and reduce their impacts in real-world setup.}
\vspace{-.2em}
\begin{itemize}
%\vspace{-.5em}
  %  \item In a particular study, in most of the cases, all three types of classification are missing
   % \item Most of the studies considered oversampling or undersampling to overcome the imbalance dataset problem. However, they didn't give any test results after the model's training based on the original dataset (test portion). So, their training and test data was completely selected from the over or undersampled dataset.
    %\item Improving accuracy metrics for the minority classes.
    %\item Providing feature-based explanation in global and local scope using explainable AI (XAI) methods (e.g., SHAP) for intrusion detection in IoT environments. 
    \item Propose a feature explanation-based step-wise identification of false-positive and false-negative intrusion instances by visually analyzing feature explanations SHAP plots when overlapped with true-positive and true-negative group explanations. %  for the {\em false positive (FP)}, {\em false negative (FN)}, {\em true positive (TP)}, and {\em true negative (TN)} groups. 
    \item Provide case study with multiple real-world intrusion datasets to showcase the efficacy in reducing misclassification (e.g., FPs and FNs) for eliable decision-making. % for network and IoT intrusion detection scenarios.  
\end{itemize}

\noindent Paper organization: Section \ref{sec:related-work} presents the related works. Section \ref{sec:method} presents the research questions and research methodology. Section \ref{sec:case_study} depicts the case study results on various network traffic datasets. Section \ref{sec:discuss} discusses the limitations of the paper while section \ref{sec:conclusion} concludes the paper.

%% file: related-work.tex
\vspace{-1mm}
%\subsection{Internet of Things (IoT) Intrusion Detection Systems}
%. Researchers have long since tried to come up with diverse solutions to automate the task. 
In the literature of intrusion detection system (IDS), we find rule-based \cite{nimbalkar2021analysis,ferrag2020rdtids,kumar2021uids}, machine learning (ML) based \cite{ml_dl_ids,islam2021towards,saheed2022machine,verma2020machine,Khan2024}, deep learning (DL) based \cite{ge2019deep,khan2021deep,electronics13061053_2024,Jony_Arnob_2024}, and hybrid \cite{hybrid_ids,sadikin2020zigbee,saif2022hiids} detection approaches. Many of these studies have used labeled intrusion dataset from the Canadian Institution for Cybersecurity (CIC), such as CIC-IDS2017 \cite{CIC-IDS-2017}, CSE-CIC-IDS2018 \cite{CIC_18}, CIC-IoT2022 \cite{CIC22}, and CICIoT2023 \cite{CICIoT-2023-datasetpaper}. Additionally, in recent years researchers have also leveraged generative adversarial network (GAN) models and gained comparatively higher accuracy in IDS with imbalanced data \cite{gan_ids_paper}. %In literature, we find some recent works using the CIC-IoT 2023 dataset \cite{} for IoT intrusion detection with tree based ML classifiers, deep learning models, and clustering approaches. 
However, very few of the existing studies focused on either the XAI approaches or focus on the reduction of false-positives and false-negatives. The primary issue with most of existing research is the black-box nature of using machine learning or deep learning models where the individual detection explanation are not provided for reliable decision-making from the model outcomes \cite{XAI_Decision}. We have further investigated some recent developments that leverage explainable AI (XAI) approaches such as SHAP and LIME to provide both global and local feature explanations for explainable intrusion detection \cite{IDS_XAI_2,IDS_XAI_3,XAI_IDS_mahmoud23,kalakoti2024improving,exai_ieee_access_2024,OmarAbdel_2023_XAI_IoT_Intrusion}. %\cite{IDS_XAI_2,IDS_XAI_3,IDS_XAI_4,IDS_XAI_5,IDS_XAI_6,XAI_IDS_mahmoud23,kalakoti2024improving,exai_ieee_access_2024,OmarAbdel_2023_XAI_IoT_Intrusion}. 
%with improved classification accuracy and transparency. Additionally, Osvaldo \textit{et al.} \cite{} discussed on evaluating various black-box XAI models for effectiveness within network intrusion detection systems. Arisdakessia \textit{et al.} \cite{} presented a survey on IoT intrusion detection with recent developments such as federated learning, explainable AI, game theory, and social psychology. 
In other studies, researchers have proposed XAI for enhancing anomaly detection in IoT and health care monitoring systems \cite{2024IEEE_XAI-IoT,Abououf2024ExplainableAF}. However, there is very limited empirical and methodological studies to showcase the usage of XAI for reducing misclassification in ML-based intrusion detection. Among such works, Lopes \textit{et al.} \cite{IDS_FP_reduce} proposed to reduce false positive identification through a secondary ML model trained with XAI attributes and Kim \textit{et al.} \cite{Kim2021Cost} proposed FOS (feature outlier score) threshold that takes into consideration the relation between an specific instance's SHAP value with another similar instance's SHAP values in terms of mean and standard deviation. Nonetheless, there are several drawbacks with this approaches, such as, no measurements for false negative detection, manually find and choose the FOS threshold, testing done on only one datasets, and decrease in TPR (true-positive-rates). Another recent study highlights a binary classification on metabolomics datasets \cite{XAI_HEALTH_FP_FN_SHAP}, which provides simple \textit{`TP vs FP'} and \textit{`TN vs FN'} scenarios based on SHAP values and local waterfall plots but did not provide any systematic process for correctly identifying FP and FN cases. Moreover, Wei \textit{et al.} \cite{xNIDS_2023_usenix} introduced {\em xNIDS}, a deep learning based model with potential FP reduction-ability but lacks global interpretability and create high sparsity of the explanation to lose vital information. % analysis (high sparsity of the explanation) can sometimes lose vital information which are necessary for the analyst. 
Furthermore, Yang \textit{et al.} \cite{CADE_2021_usenix} proposed {\em CADE}, an unsupervised DL explanation to detect concept drift, which provides better performance in malware detection, but optimal in IDS scenario. Lastly, Han \textit{et al. } proposed {\em DeepAID} \cite{DeepAID} that performs poorly when there are feature dependencies in the dataset which is often the case in IDS. In summary, all the existing studies on identifying mis-classification have not provided any systematic approach that can guide an analyst to make trustworthy decisions.

%% file: methodology.tex
\vspace{-0.4em}
\subsection{Research Questions}
This paper addresses the following research questions.

\ignore{{\color{red}\noindent \textbf{RQ1:} How can we solve the imbalance dataset problem so that the prediction results get reflected on the actual dataset yet with higher accuracy? 
}\footnote{I do not think this RQ is novel and also what is the importance in the context of FP or FN reduction?} {\color{blue} This RQ was removed and I put a new RQ3.}}

\noindent \textbf{RQ1:} Are the feature explanations for overall false-positive (FP) group differ significantly from the true-positive (TP) group? % and true negatives, respectively to identify them effectively? 

\noindent \textbf{RQ2:} Are the feature-based explanations for false-negative (FN) group differ from those of true-negative (TN) group? %and true negatives, respectively to identify them effectively? 

%{\color{red}\noindent \textbf{RQ2:} Can local explanation within XAI help in the scenario of any individual attack detection to provide more transparency for IDS? }\footnote{also this RQ is non significant as XAI supposed to give transparency as per existing research.....Let's think of re-writing the significant RQ in the context of the present problem and the solutiuon space that we are proposing! {\color{blue} Actually, sir, I am not able to find other RQs than these tows right now although RQ1 and RQ3 are very similar.}}

\noindent \textbf{RQ3:} Given local SHAP feature explanation plots are leveraged for any traffic classification, what methodological steps are required to aid the analysts potentially identify if individual traffic instances are false positives (FPs) or false negatives (FNs) identification? % prediction from the assistance of XAI?

\noindent \textbf{RQ4:} Can analysts reduce false positives (FPs) and false negatives (FNs) within a IDS in practical settings leveraging our visual analysis approach?

\vspace{-1.2mm}
\subsection{Problem Background and Formalization}
%\vspace{-0.5em}
Given a network traffic data, let's consider the $i$-th network traffic data point, ${\tt tf}_{i}$, consists of $n$ number of extracted features $ F_i = \{f_1, f_2, \cdots, f_n\}$ with a corresponding label $l_i$. Here, label $l_i$ can be either binary categories of {\em `attack'} versus {\em `benign'} or a multi-class categories if various attack types are labeled. %However, if the dataset has more granular attack type categorization, then \textit{`attack'} labels can further be replaced with one of the attack categories from the following attack category set \{{\em `DDoS'}, {\em `DoS'}, {\em `Brute Force'}, {\em `Spoofing'}, {\em `Recon'}, {\em `Web-based'}, {\em `Mirai'}\} or it can be further granular attack types as highlighted in Fig. \ref{fig:dataset class distro}. 
Next, we train machine learning supervised models $M_1, M_2, \cdots, M_m$ with the $n$ selected features and the labels extracted from the traffic dataset. %features on the test data for evaluating the model performance. % to predict the label for that traffic either as \textit{benign} or one of the \textit{attack} types. %{\color{blue} After train the models, we get the best performed model with expected accuracy and other evaluation metrics which provide a supportive analogy for the analyst to scrutinize the prediction outcome further with XAI yet with a previous confidence. To boost this confidence more, we also generate the \textit{Brier Score} for the respective model that allows us to know how close the model predicted raw probability values are to the actual outcomes.} 
These trained models can then be used to detect intrusions from the new incoming network traffic. To evaluate the confidence of such a detection model, we use \textit{Brier Score} ($Br_{m}$) for the $m$-th corresponding model. %and it can not be generalized for further action on the predicted outcome in a trustworthy manner. 
Next, we test the models on our unseen test dataset and based on the evaluation we select the best model. We also create the SHAP {\em Explainer} object $O_e$ with the selected model and the training data. The data indices for four subgroups- true positive ($D_{tp}$), true negative ($D_{tn}$), false positive ($D_{fp}$), and false negative ($D_{fn}$) are also created from the testing data. Moreover, we generate group-wise feature explanation with mean SHAP values for each of the subgroups, such as-- true-positive ($E_{tp\_mean}$), true-negative ($E_{tn\_mean}$), false-positive ($E_{fp\_mean}$), and false-negative ($E_{fn\_mean}$) groups mean SHAP values along with the global mean SHAP. %The bar plots provides the explicit picture of feature contribution on global scale and the summary plots indicate the top contributing features and their corresponding value-ranges for a particular class label. %\footnote{do we have that in the current experiments? {\color{blue} Yes sir. I have implemented the experiment as stated in this paragraph.}} %\footnote{what does bar and summary plot infer? describe in one sentence here}} 
In the subsequent process, we generate prediction label $Y_i$ using {\em predict} function and raw probability value $P_i$ using {\em predict\_prob} function with the selected model. We also create local feature explanation set $E_i = \{f_1:\phi_{1}, f_2:\phi_2, \cdots, f_n:\phi_n\}$ for the $i$-th traffic data using the SHAP {\em Explainer} object $O_e$. %for any individual incoming traffic data, ${\tt tf}_{i}$'s, label prediction 
Here $\phi_n$ indicates the importance factor (e.g., SHAP contributions) of the corresponding feature $f_n$. The local explanation for $i$-th instance, $E_i$ can be visualized as local SHAP bar plots overlapped with the group-wise (i.e., TP, TN, FP, FN) bar plots for further analysis by human analysts. %This plot helps to identify the difference of individual instance's feature contributions from the group-wise plots. 
The SHAP {\em plot-similarity} based visual analysis can be conducted by observing the overlapping bar counts for the individual instance with the specific sub-groups. We hypothesize that this proposed plot-similarity based method can effectively indicate whether the individual instance predicted as {\em attack} (checked with TP and FP groups) or {\em benign} (checked with TN and FN groups) is a correct prediction or not. %is actually  true , is closer to the TP or FP group's explanation. Similarly, if the model predict an individual instance as {\em benign}, then the visual analysis of explanation plots can give reliable indication if the identification is closer to the TN or FN group's explanation. 
Additionally, we consider the raw probability {\em $P_i$} alternatively for the model findings if visual {\em plot-similarity} method does not provide clear decisions. The algorithm process is presented in Algorithm \ref{alg:shap_algo} (see Appendix). %outline of the process is given through Algorithm . % of the step-wise methodology and the case study with multiple real-world network traffic datasets. % a true    Next, we propose to visualize this feature explanation set We propose that visual analysis of local explanation method along with the numerical presentation of the top features' contribution and previous intuitions about the actual numeric value ranges for those corresponding features %\footnote{how do we know previous value ranges? are we saving it somewhere in the process? {\color{blue} Sir, a comprehensive statistics of the dataset is given in the dataset paper, so when we have the top contributing features in our hand, we inherently have the basic statistics for them. This information is needed, cause even with the SHAP visuals, sometimes the misclassification can not be totally eradicated by the SHAP plots only, so by checking the value we can be sure. Like. But, we have to decide what is the high, low and avg value for a particular features.}} 

%can also be used for reducing false-positive (FP) and false-negative (FN) classification of the IDS. To identify FPs and FNs in case of binary classification scenario, we first find the top $x$ features (out of the $n$ features) for true-positive (TP) and true-negative (TN) groups and then compare each individual prediction's feature contributions for those $x$ features to analyze how close it is to the true-positive group (if the individual prediction is \textit{`attack'}) or true-negative groups (if the individual prediction is \textit{`benign'}). Our hypothesis behind this intuition is that the false-positive (FPs) group in general will have a different set of top features than the true-positive group and the same scenario between feature contributions for false-negative versus true-negative group. This will give analyst an additional information before making any decision with the predicted outcomes to enhance the reliability and trustworthiness.      %\footnote{please add the brier score in the formalization context, where and how can we use it and for what purpose? {\color{blue} Added}}

%\noindent \textbf{RQ4:} Can XAI help to reduce the FPR or FNR of the intrusion detection system?
\begin{figure*}[!t]
    \centering
    \includegraphics[width=0.82\textwidth, height=0.29\textwidth]{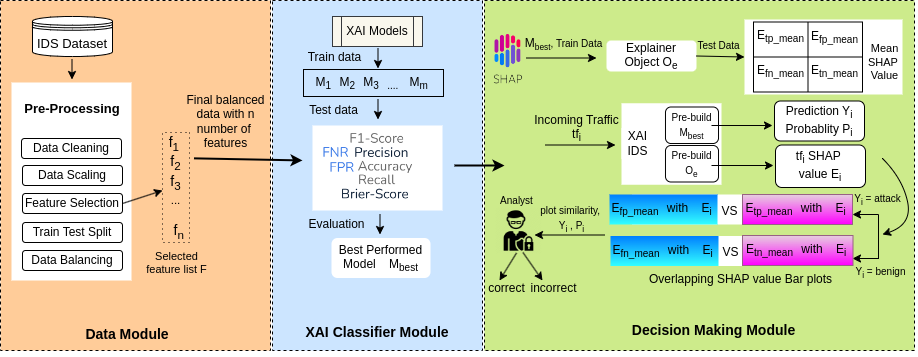}
    \caption{Overview of the proposed methodology}
\label{fig:overview}
\vspace{-1.5mm}
\end{figure*}

\vspace{-2mm}
\subsection{Overview of Methods}
\vspace{-0.4em}
In order to answer the above RQs, we have proposed the following \textbf{three} modules for our approach: (i) Dataset collection and pre-processing; %(ii) Features selection;}\footnote{are we reducing features or all features are considered?} %(iii) Balance the Training dataset 
(ii) Training, testing and evaluation of supervised XAI models; (iii) Visual analysis of feature-explanation plots to identify FPs and FNs. % Feature-based Local Explanations for False Positive and False Negative Groups. 
Figure \ref{fig:overview} further illustrates our proposed methodology. %We have further discussed each of these modules below to present the case study design.

%\footnote{overview figure-1 looks so blur, please take care of the image quality and the test size should be little smaller....it's currently bigger that the regular text font size}
\subsubsection{Dataset Collection and Pre-Processing}
We rely on any traditional IDS or IoT network traffic dataset for applying our methodology. %Our methodological approach can be applied to any IoT and other network traffic based dataset where we have planned to select the recently published dataset by the Canadian Institute for Cybersecurity named CIC-IoT-2023 \cite{s23135941}. %published by  for this study. %, which is intended to help develop security analytics tools for the IoT context. 
Any relevant traffic dataset $D$ would contain various labeled cyber attacks as ground-truths. %This dataset includes %46,686,579 instances from which 
%46,686,545 unique rows 
% \footnote{what are these connections? I am thinking of this many rows. {\color{green}check done.}} 
Generally the intrusion detection traffic dataset contains the following high-level attack categories: {\em DDoS}, {\em Denial of Service (DoS)}, {\em Recon}, {\em Web-based}, {\em Brute-Force}, {\em Spoofing}, {\em Mirai}, {\em Infilteration}, {\em Exploits}, {\em Fuzzers}, {\em Backdoor}, {\em MITM}, {\em Ransomware}, {\em Shellcode} and {\em Worms} %\footnote{we need to add more attacks if other dataset has other attack labels}. \
However, in this paper we would focus on the binary classification scenario where regular traffic are labeled as `{\em benign}' and various attack traffic are labeled as `{\em attack}'. % tasks of malicious versus benign groups. 

\noindent \textbf{Handling Imbalance Dataset.} %In most of the existing studies, when data balancing has been conducted, the testing portion is also impacted with the balancing, which may create duplicate entries in the testing for oversampling methods. Thus, %researchers used any balancing methods like SMOTE {(\tt S)} or Random Undersampling {(\tt U)}, they took their testing portion of the data after they applied the balancing method which resulting in many duplicate entries in the testing data. Because it has been proved that using the original set of testing portions can result in lower accuracy value \cite{Wang2021} but may provide a more realistic performance of the model. Thus, 
For data imbalance problem, we first split the entire dataset into 80:20 for training and testing, respectively. Then, we apply the oversampling (e.g., SMOTE \cite{Khanday_Fatima_Rakesh_2023}), undersampling (e.g., Random Undersampling) or a combination of both only on the training portion of the data, which leads to more realistic and reliable performance results. Also, to get fair result with the test data, we have balanced test data by applying random undersampling to avoid any redundant encounter on data class which is comprised of equal number of attack and benign samples. %\footnote{is this paragraph still valid with the present experiments? {\color{blue} Yes sir. I have implemented over or under sampling in each dataset cases.}}%\footnote{why not Random undersampling on the testing portion, which would balance the testing as well as solve the issue of duplicates} %\footnote{when you mention S+U it seems smote and undersampling is coupled together, how exactly is this performed is not clearly highlighted ...Is it the case where some classes are oversampled and others are undersampled and then the final dataset is complied? The table 4 needs more context of the comparison ...to be read clearly!} %, we do it only on the training portion of the data. 
%Hence, we believe our test results reflect more credible results on the original dataset on the training portion of each classification tasks. So, we've basically applied the SMOTE for the minority classes and Random Undersampling for the majority classes.

%\footnote{if all features are selected, then what is the purpose of this sub-section? {\color{blue} Sir, we have considered all the features, but before that we have tested with reduced feature list with 24 features, (see table 3). In that we got a comparitively lower accuracy, so for this dataset, we consider the whole feature list, but for another dataset ,reduced feature lsit can be effective}}

\subsubsection{Training, Testing and Evaluation of Supervised XAI Classification Models}
We use SHAP as an XAI module that works better with tree based classifiers \cite{linardatos2020explainable}. In this paper, we use several tree-based classifiers: Decision Tree (DT) \cite{dt_paper}, XGBoost (XGB) \cite{xgb_paper}, and Random Forrest (RF) \cite{rf_paper} for the classification tasks. %(i.e., binary, group, and individual classifications). 
For all the models, the training and testing would be conducted on balanced dataset. %while the testing has been conducted on the original data portions splitted for test before the balancing. 

%Our main target is to build the classifier in such a way that it can generate relevant results for testing data however with less complexity and overhead. Also, tree-based models have been proven to work very well with SHAP \cite{fast_shap}. 

%\vspace{-0.9em}
% \subsubsection{Evaluations of XAI Models}
%\footnote{also mention about the confusion matrix.. {\color{blue} done}}
For all the classification models, we consider the following standard performance metrics %\cite{abdelmoumin2021performance}
: {\em accuracy}, %table \ref{table:comparison}). 
{\em precision}, {\em recall}, {\em average F1-score}, and {\em Brier score (BS)}. The {\em Brier score (BS)} is defined as:
 %  are taken as the weighted average over all the classes. %The equations for accuracy, precision, recall, and F1-score are as follows:
\ignore{
\begin{align*}
    Accuracy & = \frac{TP + TN}{TP + TN + FP + FN} \\
    Precision & = \frac{TP}{TP + FP} \\
    Recall & = \frac{TP}{TP + FN} \\
    F1\text{-}score & = 2 \times \frac{Precision \times Recall}{Precision + Recall}
\end{align*}

where:
\begin{align*}
    TP & : \text{True Positives} \\
    TN & : \text{True Negatives} \\
    FP & : \text{False Positives} \\
    FN & : \text{False Negatives}
\end{align*}
}
%We also considered the the Brier Score(B.Scr.) to further give justification to our model's output because in all of the found previous studies, the explanation artificial intelligence (XAI) or other kind of interpretability is missing. 
$
    BS = \frac{1}{N} \sum_{i=1}^{N} (fr_i - o_i)^2
$%\end{equation}$ 
where, %$BS$ presents the {\em Brier Score}, 
$N$ resembles the number of samples, $fr_i$ represents the forecast probability of an event for $i$-th sample, and $o_i$ represents the observation of an event for $i$-th sample. To contextualize the {\em Brier score} as an evaluation metric, a lower score (closer to zero value) indicates that the model is very confident about the output it generates as the numerical probabilistic difference between the predicted output (the raw output generated by the model before taking the softmax) and the ground truth \cite{rufibach2010use_brier}. Moreover, we review the confusion matrix to assess the ratio of correct (TP, TN) and incorrect (FP, FN) predictions from the test data. Additionally, we evaluate our approach with other existing literature to justify the competitiveness of the proposed approach. % \footnote{need to move current section 4.1 to here ...also the brier score part {\color{green}done}}

%\vspace{-0.6em}

%\footnote{please address Line-19 of algorithm to formalize it algorithmically {\color{blue}, Sir, if we want to make it more mathemetical linguistic, we need to introduce more notaions so, for now I can't think of how to put it in a more matehmetical way with the current context also with minimal notations.}}

\subsubsection{Visual Analysis of Explanation Plots For FPs and FNs Identification}
\label{sec:visual_analysis_steps}
%In this work, we use SHAP as a means to explain our ML models, which prove to be a powerful tool in XAI. %that uses game theory to measure player contribution to a certain move or outcome. 
SHAP, a popular XAI module, incorporates different visualization plots which provide feature-wise explanations for the whole model, selective group of instances, or any individual instances. %that can be very crucial in visualizing the feature importance for a particular ML model, any specific group, and any single instance. 
In our proposed approach, we try to generate feature-based SHAP explanation bar plots using the SHAP {\em TreeExplainer} for true-positive (TP), true-negative (TN), false-positive (FP), and false-negative (FN) groups within the model. Now, when an analyst use our proposed intrusion detection system in practice with the local explanation enabled for any individual traffic instance prediction, they can conduct the following three-steps (S1--S3) process to reach a reliable and trustworthy decision-making: %\footnote{explicitly mention what items to look for here}

    %\item \textbf{Getting the model performance evaluation and confidence score beforehand.}
\noindent \textbf{S1 (L1-L7 in Algorithm \ref{alg:shap_algo}):} Generating and storing the top contributing features' (usually top 20 features) %if \#all\_feature < 60 else 30) 
    SHAP bar plots with global mean SHAP values for all four groups (e.g., TP, TN, FP, and FN). %\footnote{by observing the bar graph, right? {\color{blue} added}}
    % \item Identify at least the top $\sqrt{(Num\_of\_total\_features)}$ number of appeared positive and negatively contributing features among the selected top features from SHAP bar plot %\footnote{can we come with specific schemes? or what can be a good rationale here? {\color{blue}min(sqrt{feature\_number},total appearance within top 20/selected features), I found it in a blog that I can't remember at this moment but we can't cite it, here, so maybe you can describe it better, sqrt(feature\_number) for each fp, fn, tp, tn groups including both positive and negative SHAP values}} 
    % for all four (TP, TN, FP, and FN) groups. For example, if we have 49 total features then approximately, $\sqrt{49}$ = 7 features can be analyzed for each positive contributing and negative contributing cases within the top 20-30 contributing features. 
    
    % \item Having the overview of the value ranges (high, medium, or low) in the summary plot of those features that differ in TP vs FP and TN vs FN scenarios, their importance on a particular prediction class. %from the summary plot. %\footnote{all of the top 20 value range? how to know the range, where to cut off? be specific... {\color{blue} for a specific case, like in TP, sqrt(feature\_number) is good enough}}
    
\noindent \textbf{S2 (L8-L15 and L22-L25 in Algorithm \ref{alg:shap_algo}):} For each individual instance outcome, if the prediction is {\em positive} (meaning an attack traffic is predicted), then generate plots using the local feature SHAP value $E_i$ by comparing the global SHAP values of the true-positive ($E_{tp\_mean}$) and false-positive ($E_{fp\_mean}$) group's top features through a new overlapping bar graph. On the other hand, if the prediction is {\em negative} (meaning a benign traffic is predicted), then the local features' SHAP values would be mapped in overlapping bar graphs with the corresponding features from both the true-negative ($E_{tn\_mean}$) and false-negative ($E_{fn\_mean}$) groups. %This comparison is depicted by two overlapping bar plots, i) instance's SHAP value and either TP or TN mean SHAP value ii) instance's SHAP value and either FP or FN mean SHAP value.}

\noindent \textbf{S3 (L16-L20 and L26-L30 in Algorithm \ref{alg:shap_algo}):} In this step, we observe the overlapping graphs to understand visually differentiable or similar feature contributions. We can infer that higher number of overlapping bars in these bar graphs which we define as $plot\_sim$, indicates a particular instance is closer to that corresponding group while the less overlapping scenario indicates distance from that group. Using this metrics from the respective graphs, an analyst can finally take the decision to mark a prediction as correct (TP, TN) or incorrect (FP, FN).

% In case where clear visual differences are evident, we observe the number of direction altering features, meaning how many features with positive SHAP values in a group explanation bar graph, but the local instances have negative SHAP values for those features and vice versa.  %particular features the group-wise SHAP values and the individual instance's SHAP value are in opposite direction, meaning if for feature $f_x$ the SHAP value is positive in a FP group bar plot   %We would also check if a high number of features have altering directions, meaning the indivu %so by visually assessing the overlapping SHAP bar plots an analyst can identify whether the prediction is correct.

%\item S4: %\footnote{so we are not addressing it in the algorithm, correct? this is a optional manual observation step if we can not determine from S1-S3.{\color{blue}The S3 is the decision making process so, if we can not detrmine after that, we can consider this option.}}

\ignore{This step is optional and only applicable in cases where the decision of similarity or dissimilarity is not clear in S3.} 
There can be instances where the overlapping bar plots are not clearly giving the analyst a clear hint for reliable decision-making. For example, a positive prediction of an individual traffic instance may have very similar overlapping graphs with both TP and FP groups, then the analyst can rely on model's original prediction $Y_i$. %A higher score on these highlighted measurements are also an indicator of correct prediction. 

%% file: experiment-results.tex
%\footnote{reviewer comment 3: only one dataset was used in the evaluation, it remains unclear whether it is effective.}
%\footnote{Reviewer comment: Different datasets might have varying distributions, attack types, and feature sets, which could affect the performance and reliability of the proposed approach. (same as previous comment)}
%\footnote{reviewer comment: the paper mentions that the training data is balanced, it is unclear if any balancing was done on the testing data. Evaluating the model on imbalanced test data can still lead to biased performance metrics, particularly for minority classes. The model might appear to perform well overall, but its performance on minority classes might still be inadequate.}

\vspace{-1.3mm}
\subsection{Dataset Highlights and Data Pre-Processing}
\vspace{-1mm}
We have conducted our approach on three different network (including IoT network) traffic datasets. The main goal is to test and show if the proposed methodology is applicable in various dataset scenarios. %strengthen our research objective by showing that out proposed method can be adopted on different use case scenarios. 
Hence, we first provide a brief highlights of the three differently sourced datasets used here.
% in our case study. %Some information about the used datasets and their pre-processing steps are discussed below:

\begin{table*}[!ht]
    \centering
    \caption{Brief Details of Intrusion Datasets for Case Study}
     \resizebox{0.70\textwidth}{!}{\begin{tabular}{cccc|ccc|cc|cc}
        \hline
        \multirow{2}{*}{\textbf{Dataset}} & \multirow{2}{*}{\textbf{Published Year}} & \multirow{2}{*}{\textbf{\# Total Rows}} & \multirow{2}{*}{\textbf{\# Features}} & \multirow{2}{*}{\textbf{Category}} & \multirow{2}{*}{\textbf{\# Type}} & \multirow{2}{*}{\textbf{Count}} & 
        \multicolumn{2}{c}{\textbf{\# Initial Split (80:20)}} & \multicolumn{2}{c}{\textbf{\# Balanced Class}}\\
        \cline{8-11}
         & & & & & & & \textbf{Train} & \textbf{Test} & \textbf{Train} & \textbf{Test}\\
        \hline
        \multirow{2}{*}{$D_1$} & \multirow{2}{*}{2023} & \multirow{2}{*}{46,686,579} & \multirow{2}{*}{46} & Benign & 1 & 1,098,195 & 878,447 & 219,748 & 1,000,000 & 219,748\\
        & & & & Attack & 7 & 45,588,384 &  36,470,816 & 9,117,568 & 2,000,000 & 219,748\\
        \hline
        \multirow{2}{*}{$D_2$} & \multirow{2}{*}{2021} & \multirow{2}{*}{4,276,736} & \multirow{2}{*}{41} & Benign & 1 & 1,415,361 & 1,132,221 & 283,140 & 1,132,221 & 283,140\\
         & & & & Attack & 20 & 2,861,375 & 2,289,167 & 572,208 & 1,132,221 & 283,140\\
        \hline
        \multirow{2}{*}{$D_3$} & \multirow{2}{*}{2021} & \multirow{2}{*}{555,278} & \multirow{2}{*}{81} & Benign & 2 & 517,582 & 414,022  & 103,560 & 30,200 & 7,496\\
         & &  &  & Attack & 4 & 37,696 & 30,200 & 7,496 & 30,200 & 7,496\\
        \hline
    \end{tabular}}
    \label{tab:dataset_table}
\end{table*}

\begin{table*}[!ht]
    \centering
    \caption{Performance Evaluation of Various ML Models with Different Intrusion Detection Datasets}
     \resizebox{0.50\textwidth}{!}{\begin{tabular}{c|cccccccc}
        \hline
        \textbf{Dataset} & \textbf{Model} & \textbf{FPR} & \textbf{FNR} & \textbf{Precision} & \textbf{Recall} & \textbf{F1-Score} & \textbf{Accuracy} & \textbf{Brier Score} \\
        \hline
        \multirow{3}{*}{$D_1$} & DT & 0.04 & 1.56 & 99.21 & 99.20 & 99.20 & 99.20 & 0.0079 \\
         & XGB & 0.12 & 0.51 & 99.68 & 99.69 & 99.68 & 99.68 & 0.0028 \\
         & RF & 0.05 & 2.01 & 98.97 & 98.99 & 98.97 & 98.97 & 0.0082 \\
        \hline
        \multirow{3}{*}{$D_2$} & DT & 2.85 & 5.69 & 95.77 & 95.73 & 95.73 & 95.73 & 0.0310 \\
         & XGB & 0.50 & 1.43 & 99.04 & 99.03 & 99.03 & 99.03 & 0.0080 \\
         & RF & 0.72 & 7.25 & 96.21 & 96.02 & 96.02 & 96.02 & 0.0380 \\
        \hline
        \multirow{3}{*}{$D_3$} & DT & 16.97 & 0.04 & 91.50 & 92.72 & 91.50 & 91.50 & 0.0726\\
         & XGB & 14.10 & 0.16 & 92.87 & 93.72 & 92.86 & 92.83 & 0.0614 \\
         & RF & 16.25 & 0.01 & 91.87 & 93.00 & 91.87 & 91.81 & 0.0708 \\
        \hline
    \end{tabular}}
    \label{tab:performance_table}
\end{table*}

\noindent \textbf{CIC-IoT-2023 dataset ($D_1$):} The Canadian Institute for Cybersecurity published CIC-IoT-2023 dataset \cite{CICIoT-2023-datasetpaper} that includes 46,68,6579 instances in total, with 45,58,8384 (97.65\%) being attack instances and 10,98,195 (2.35\%) being benign instances containing 46 features (columns). This dataset provides a total of 7 high-level attack types-- {\em DDoS}, {\em Denial of Service (DoS)}, {\em Recon}, {\em Web-based}, {\em Brute-Force}, {\em Spoofing} and {\em Mirai} those are further re-categorized in 33 individual attack sub-types. As it is a highly imbalanced dataset, we have done some pre-processing before using it for our case study. First, we drop all the null values from the dataset. Next, we split the dataset into $80:20$ for training and testing. This ratio is adopted from the previous studies on the same dataset \cite{CICIoT-2023-datasetpaper, Khan2024}. Next, we apply SMOTE on the majority class ({\em attack} class) and random under sampling %using the {\em Imblearn} library 
on the minority class ({\em benign} class) for the training portion. However, to get fair results on test data, we have down-sampled randomly only the majority class to generate equal number of {\em attack} and {\em benign} cases in the test data. In both train and testing section, all kinds of attack types are present. A brief dataset detail is presented in Table \ref{tab:dataset_table}.

\noindent \textbf{NF-UQ-NIDS-v2 dataset ($D_2$):} The University of Queensland- Australia has published the extensive network intrusion detection system dataset known as NF-UQ-NIDSv2 \cite{sarhan2023nf}. This dataset has a total of 75,987,976 records, where 25,165,295 (33.12\%) are {\em benign} records and 50,822,681 (66.88\%) are {\em attack} instances. However, due to space and computational constraints, we have used a random portion of the dataset which left us with a total of 4,276,436 (5.63\%) instances where 2,861,375 (66.90\%) {\em attack} and 1,415,361 (33.10\%) {\em benign} instances are selected. This dataset contains 20 different network-based anomaly attacks including {\em Infilteration}, {\em Exploits}, {\em Fuzzers}, {\em Backdoor}, {\em MITM}, {\em Ransomware}, {\em Shellcode} and {\em Worms}. %, which was assembled using data meant to mimic actual network traffic and attack situations. 
The pre-processing steps includes firstly removing the null values. Next, we remove some feature columns such as {\em IPV4\_SRC\_ADDR} and {\em IPV4\_DST\_ADDR} as these features represent source and destination IP addresses. Then, we apply Min-Max scalar to scale down the data and split the train and test portion into standard $80:20$ ratio. Next, we apply random down-sample technique using {\em Pandas} random sampling method on both train and test data that resulted in equal number of {\em attack} and {\em benign} instances, which still preserve all the 20 attack types. A brief details of this dataset is presented in Table \ref{tab:dataset_table}. %Here also, all 20 kinds of attack types are preserved in train and test data. 

%\vspace{-1mm}
\noindent \textbf{HIKARI 2021 dataset ($D_3$):} The third dataset is HIKARI-2021 dataset \cite{ferriyan2021generating} incorporated in this study. This dataset has a total of 555,278 rows, out of which 517,582 (93.21\%) are {\em benign} instances and 37,696 (6.79\%) are {\em attack} instances. The {\em benign} instances have two sub-groups: {\em benign} and {\em background}, while the {\em attack} has four sub-groups: {\em probing}, {\em bruteforce}, {\em bruteforce-XML}, and {\em crypto-miner}. In pre-processing step, first we remove all the null value rows. Then we remove some data columns such as {\em Unnamed: 0.1}, {\em Unnamed: 0}, {\em uid}, {\em originh} and {\em responh} that have no empirical significance in classification task. Next, we proceed with a total of 81 features. Then, we split the dataset into $80:20$ train and test portion. Since this dataset is highly imbalanced, we apply the same down-sampling method that is applied on $D_2$ in order to get equal number of 
 benign and attack instances in both train and test data. A brief details of the dataset is presented in Table \ref{tab:dataset_table}. 
 %and all the class sub-types are preserved in the later balanced train and test data. 

\begin{figure*}[!t]
\centering
\subcaptionbox{$D_1$\label{fig: cm_cic}}{\includegraphics[width=0.25\textwidth,height=2.75cm]{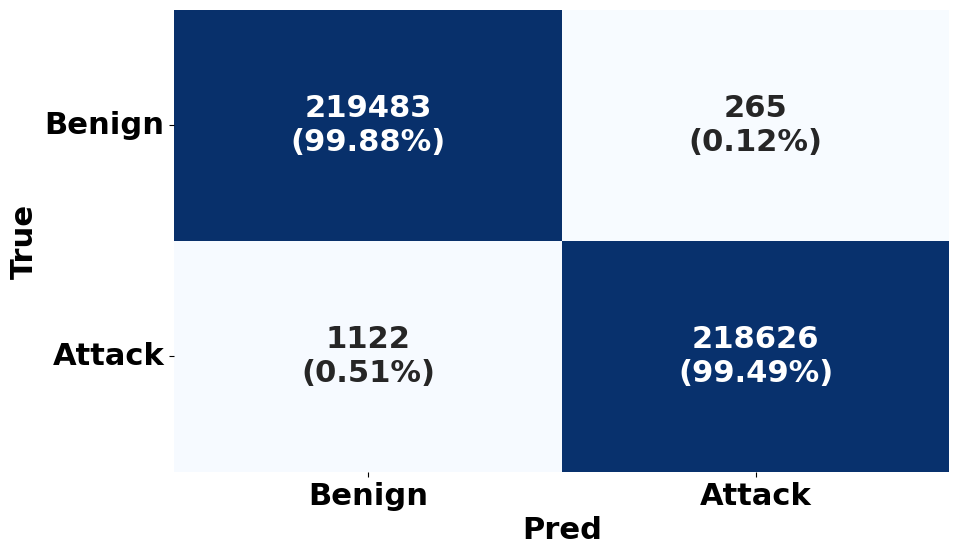}}%
\hfill % <-- Separation
\subcaptionbox{$D_2$\label{fig: cm_nf_uq}}{\includegraphics[width=0.25\textwidth,height=2.75cm]{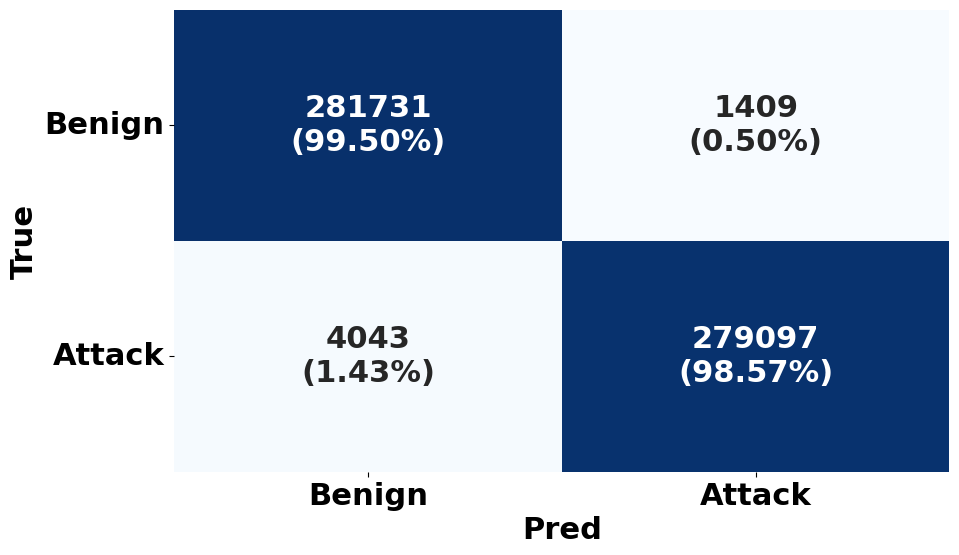}}%
\hfill % <-- Separation
\subcaptionbox{$D_3$\label{fig: cm_hikari}}{\includegraphics[width=0.25\textwidth,height=2.75cm]{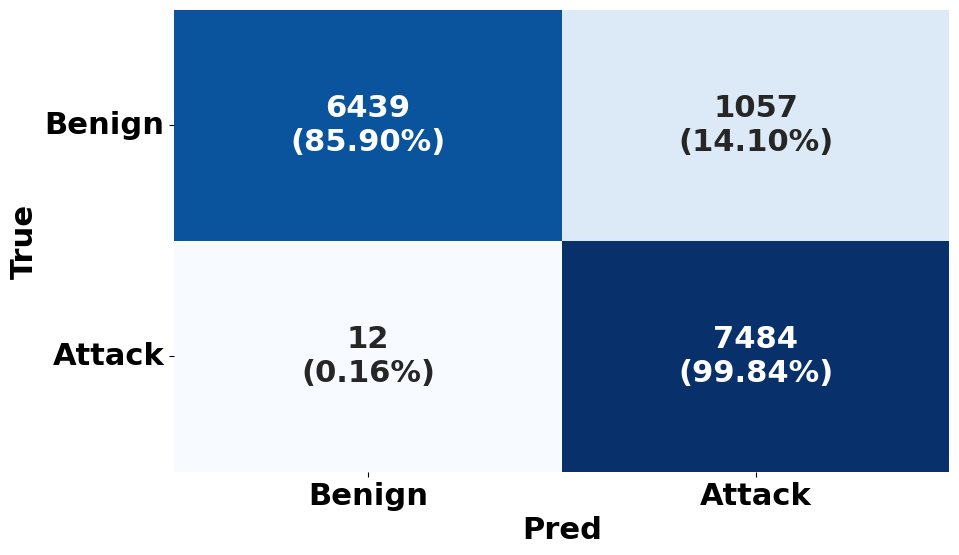}}%
\caption{Confusion matrix for the best performed XGB model}
\label{fig: cm_binary_class}
\vspace{-1.2mm}
\end{figure*}

\begin{table}[!b]
    \centering
    \vspace{-1.2mm}
    \caption{Raw Prediction Probability ($P_i$) in Percentages For All Four Cases Within The Test Data}
    \resizebox{0.38\textwidth}{!}{\begin{tabular}{c|cccccc}
        \hline
        \textbf{Dataset} & \textbf{Cases} & \textbf{$P \geq 0.70$} & \textbf{$P \geq 0.75$} & \textbf{$P \geq 0.80$} & \textbf{$P \geq 0.85$} & \textbf{$P \geq 0.90$}\\
        \hline
        \multirow{4}{*}{$D_1$} & TP & 99.94 & 99.92 & 99.90 & 99.88 & 99.85 \\
         & TN & 99.74 & 99.60 & 99.38 & 99.04 & 98.28 \\
         & FP & 36.98 & 25.66 & 19.25 & 12.83 & 9.06 \\
         & FN & 84.58 & 79.77 & 74.60 & 67.56 & 58.38 \\
        \hline

        \multirow{4}{*}{$D_2$} & TP & 99.45 & 99.23 & 98.99 & 98.30 & 97.52 \\
         & TN & 99.23 & 98.92 & 98.46 & 97.38 & 94.23 \\
         & FP & 53.87 & 44.29 & 36.41 & 28.46 & 19.80 \\
         & FN & 67.55 & 62.82 & 56.17 & 43.63 & 20.78 \\
        \hline

        \multirow{4}{*}{$D_3$} & TP & 96.95 & 92.56 & 82.39 & 66.58 & 47.82 \\
         & TN & 99.60 & 99.57 & 99.52 & 99.44 & 99.29 \\
         & FP & 92.24 & 84.96 & 68.12 & 45.88 & 29.71 \\
         & FN & 25.00 & 16.67 & 16.67 & 16.67 & 16.67 \\
        \hline
    \end{tabular}}
    \label{tab:prediction_table}
\end{table}

\vspace{-2.2mm}
\subsection{XAI Models and Evaluation}
%Based on the intuition from the previous research that amongst ML models, tree-based classifiers perform well in IDS classification task \cite{CICIoT-2023-datasetpaper, Khan2024, sarhan2021netflow, adli2023anomaly, fernandes2022network}, we have chosen three classifiers- {\em Decision Tree} (DT), {\em XGBoost} (XGB), and {\em Random Forest} (RF).
\vspace{-1mm}
We evaluate our selected tree-based classifier models with binary classification for all three datasets presented in Table \ref{tab:performance_table}. %presents the performance evaluation for the three tree-based ML models.  
It is evident that {\em XGBoost} model outperformed the other models for all the datasets in terms of standard performance metrics as well as the lower value of {\em Brier score}. We exhibit the highest model accuracy of $99.68\%$ in $D_1$, $99.03\%$ in $D_2$, and $92.83\%$ in $D_3$ for the XGB model. %We have maintained uniform parameters for the XGB classifiers for all the test experiments in each dataset. 
In this case study, we have used {\em binary:logistic} as the objective, $n\_estimators=100$ and $max\_depth=5$ as the hyper-parameters. Moreover, we observe the confusion matrix for each of the three datasets as depicted in Figure \ref{fig: cm_binary_class}(a), \ref{fig: cm_binary_class}(b), and \ref{fig: cm_binary_class}(c), respectively where we see that the accuracy for the minority class is very close to that of the majority class. However quite surprisingly, in $D_3$, the accuracy for the majority class {\em benign} is comparatively lower ($85.90\%$) that results in a higher false-positive-rate (FPR) of $14.10\%$ but low false-negative-rate (FNR) of $0.16\%$. For the other two datasets ($D_1$ and $D_2$), the FPR and FNR is reasonably lower where in $D_1$ the FPR is $0.12\%$ and FNR is $0.51\%$; in $D_2$ FPR is $0.50\%$ and FNR is $1.43\%$. 

Furthermore, Table \ref{tab:prediction_table} presents the percentage of TP, TN, FP, FN groups based on the raw probability outcomes in certain ranges, which shows the raw prediction probability is higher for the TP and TN prediction in all of the dataset cases. This provide insights that in general the models are more confident in predicting these instances and we do not want to lose too much of the TP and TN cases while correcting FP and FN cases. %finding so in parallel with the {\em Brier} score, the models are more confident on predicting these cases. This is important as we do not want to lose current TP and TN count and when a model's prediction probability value exceed a certain threshold, we can merely rely on the prediction in that scenario regardless of checking the bar plots. 
%This way we may lose some portion of FP and FN count but this won't effect much on the correct predictions. 
For instance, we can set a threshold of raw probability as a benchmark for certain datasets where we can start relying on the outcome to make a decision. The higher the probability threshold we set, the less TP and TN instances are affected, while we also need to consider improving the FPs and FNs identification. In this case study, we have set the raw emperical probability threshold to $0.90$. Additionally, table \ref{tab:comparison} presents the comparison of {\em XGB} model with our approach and other existing approaches, where it shows ours is competitive and outperforming in all cases where balanced and unique test datapoints are used. % is used anfor both classes with all unique data points. %Although the main focus of this research is to help the analyst identify the wrong prediction, we also try to make the ML models perform well enough. From table \ref{tab:comparison}, we can also see that two of the required criteria of fair test result; 1) balanced test data and 2) a unique set of test data, were not maintained in the previous studies, but we have dealt with those and provided much fairer test result compared with the other existing studies. 
% This table further highlights if the corresponding studies have used the entire dataset with SMOTE oversampling ({\tt `S'}), undersampling ({\tt `U'}), a combination of these two ({\tt `S+U'}), and if a partial portion of the dataset ({\tt `P'}) or the {\tt `full'} dataset without any sampling strategy is used for the detection study. We need the best-performed model in comparison with the existing literature studies on the resorted dataset as that gives us less incorrect predicted outcomes by the model in the first place. We also observe that the {\em XGB} model performs competitively compared to other existing works. Accuracy-wise our model performance is the best in the list among the studies that used the full dataset for $D_3$ classification and second best in the same scenario for $D_1$ and $D_2$ classification scenarios. We have also significantly improved the accuracy for the minority classes as evident from the confusion matrix (see Fig \ref{fig: cm_8}). %In most of the previous works, they claimed higher accuracy with better performance in the majority class identification and there are existing studies to report even 0\% accuracy for the minority class \cite{Jony_Arnob_2024}. 
Lastly, we observe low {\em Brier Score} (close to 0) with {\em XGB} model implying a confident model.

\begin{table}[!b]
    \centering
    \caption{Comparing Our Approach With Existing Studies}
    \resizebox{0.46\textwidth}{!}{\begin{tabular}{c|ccccc}
        \hline
        \textbf{Dataset} & \textbf{Ref.} & \textbf{Best Model} &\textbf{Acc.} &\textbf{Is Test Data Balanced?} &\textbf{All Unique Test Data?}\\
        \hline
        \multirow{4}{*}{$D_1$} & \cite{CICIoT-2023-datasetpaper} & RF & 99.68 & no & yes\\
         & \cite{electronics13061053_2024}   & LSTM & 99.99 & no & yes\\
         & \cite{Gheni2024-as} & MLP & 98.83 & yes & no\\
         & \cite{Khan2024} & RF & 99.57 & yes & no\\
         & our's & XGB & 99.68 & yes & yes\\
        \hline

        \multirow{4}{*}{$D_2$} & \cite{vishwakarma2022dids} & DNN & 98.23 & no & yes\\
         & \cite{sarhan2021netflow} & ET & 97.25 & no & yes \\
         & \cite{adli2023anomaly} & ET & 99.09 & no & yes \\
         & our's & XGB & 99.03 & yes & yes \\
        \hline

        \multirow{4}{*}{$D_3$} & \cite{ferriyan2021generating} & RF & 99.00 & no & yes\\
         & \cite{vitorino2024reliable} & LGBM & 93.20 & no & yes\\
         & \cite{fernandes2022network} & RF & 99.00 & no & yes \\
         & our's & XGB & 92.83 & yes & yes\\
        \hline

    \end{tabular}}
    \label{tab:comparison}
\end{table}

\begin{figure*}[!t]
\centering
\subcaptionbox{higher overlapping bars with FP\label{fig: fp_cic_fp}}{\includegraphics[width=0.47\textwidth,height=6cm]{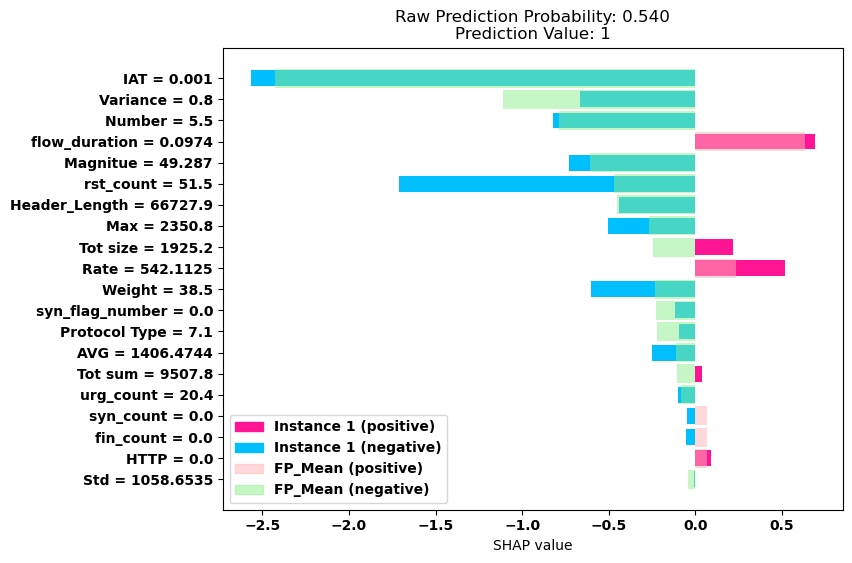}}%
\hfill % <-- Separation
\subcaptionbox{lower overlapping bars with TP\label{fig: fp_cic_tp}}{\includegraphics[width=0.47\textwidth,height=6cm]{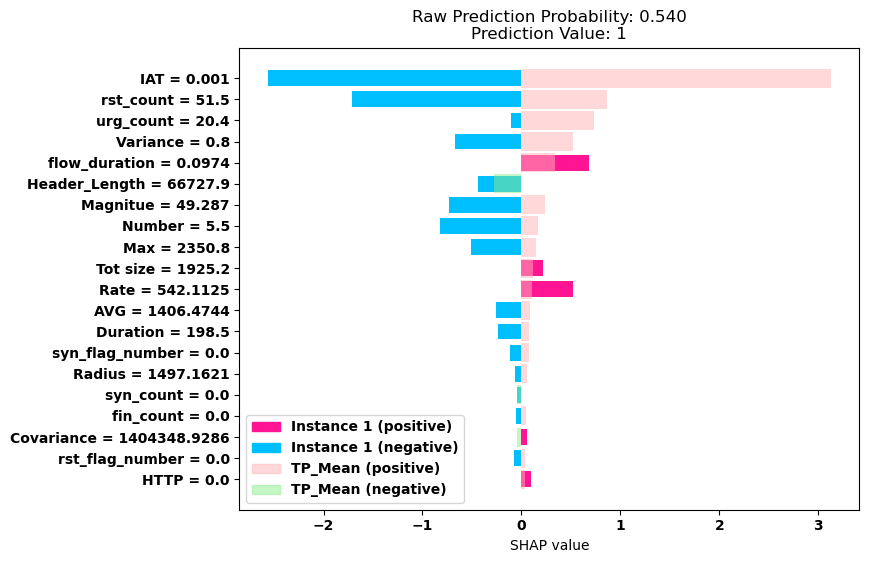}}%
\caption{Overlapping bar plots with a random false-positive instance within dataset $D_1$}
\label{fig: Dataset_1_SHAP}
\end{figure*}

\vspace{-2mm}
\subsection{False Positives and False Negatives Identification By Analyzing SHAP Plots}

\begin{figure*}[!t]
\centering
\subcaptionbox{higher overlapping bars with FN\label{fig: fn_nf_fn}}{\includegraphics[width=0.47\textwidth,height=6cm]{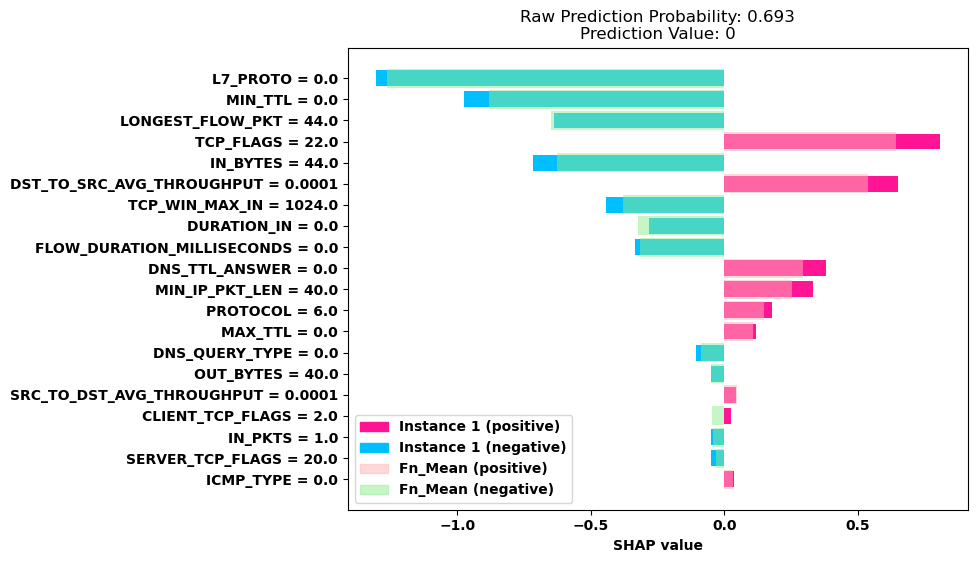}}%
\hfill % <-- Separation
\subcaptionbox{lower overlapping bars with TN\label{fig: fn_nf_tn}}{\includegraphics[width=0.47\textwidth,height=6cm]{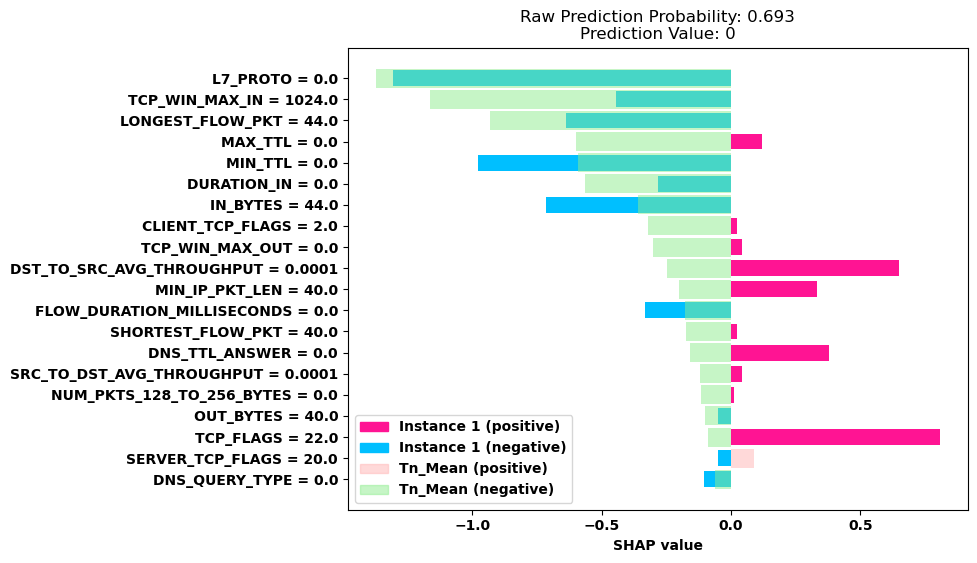}}%
\caption{Overlapping bar plots with a random false-negative instance within dataset $D_2$}
\label{fig: Dataset_2_SHAP}
\vspace{-1.5mm}
\end{figure*}

\vspace{-1mm}
\subsubsection{Generating SHAP Plots}
\label{sec:RQ_3_steps}
For each case study datasets, we apply the SHAP's \emph{TreeExplainer} object on the best-performed XGB model. Then, we have taken at most $10,0000$ random instances from the test data for TP and TN groups, respectively considering the computational cost of calculating SHAP values. Similarly, we have also taken random samples of at most $1,000$ instances for the FP and FN groups, respectively. Next, we generate the SHAP group-wise bar plots and mean SHAP values for all the four groups (TP, TN, FP, FN) where different features are in the top position of the bar plots for these different groups. %  the instances, we create SHAP values for all four cases and generate SHAP global bar plots.

\begin{figure*}[!t]
\centering
\subcaptionbox{lower overlapping bars with FN\label{fig: tn_NF_fn}}{\includegraphics[width=0.47\textwidth,height=6cm]{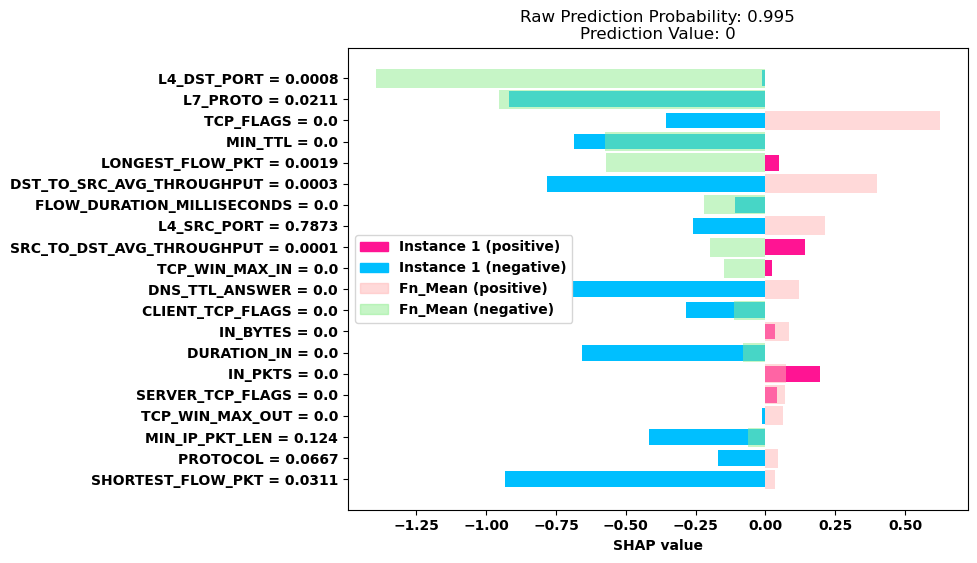}}%
\hfill % <-- Separation
\subcaptionbox{higher overlapping bars with TN\label{fig: tn_NF_tn}}{\includegraphics[width=0.47\textwidth,height=6cm]{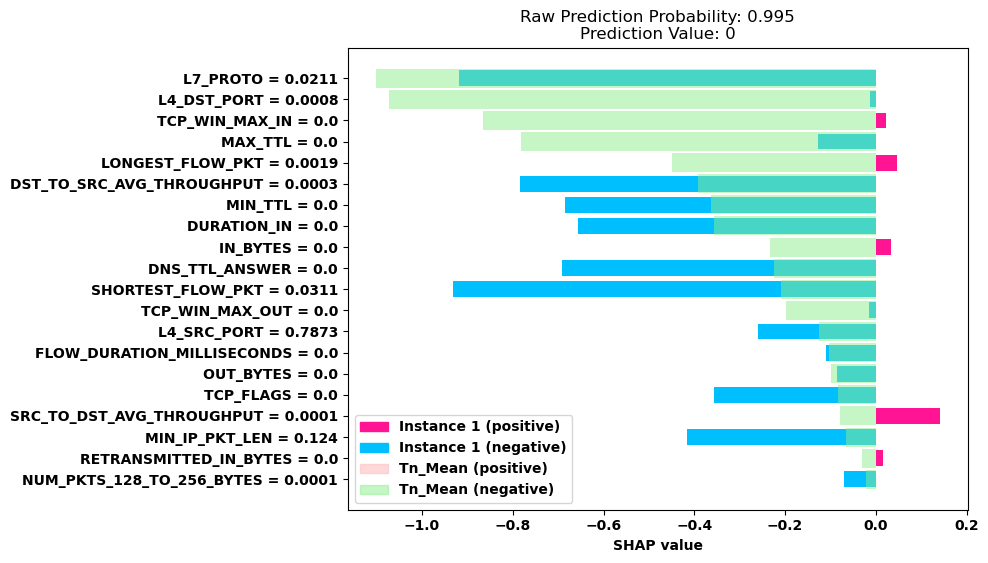}}%
\caption{Overlapping bar plots with a random true-negative instance within dataset $D_2$}
\label{fig: Dataset_2_SHAP_2}
\vspace{-1mm}
\end{figure*}

The SHAP bar plot in general provides the average Shapely value for the particular features. We can identify the top 20 most contributing features for each cases (FP, FN, TP, or TN) and save them for later comparison. Now, for an incoming traffic instance, the model makes the label prediction and generates the raw probability along with a local explanation bar plot, which can be mapped to initiate a overlapping bar graph following the process described in section \ref{sec:visual_analysis_steps}. %Then we also generate the SHAP value for that individual traffic with the help of pre-build {\em explainer} object from the corresponding model. 
Now, if the prediction is positive (i.e., {\em attack}), we generate two overlapping SHAP bar plots-- (i) the TP group average SHAP values for the top 20 TP features and the individual instance's SHAP values for the same corresponding TP features; (ii) the FP SHAP values for the top 20 FP features and the individual instance's SHAP values for the corresponding FP features. On the other hand if the prediction is negative (i.e., {\em benign}), then we generate the following two overlapping SHAP bar plots-- (i) the TN group average SHAP values for the top 20 TN features and the individual instance's SHAP values for the same corresponding TN features; (ii) the FN SHAP values for the top 20 FN features and the individual instance's SHAP values for the corresponding FN features.
%\footnote{I think we need to show the default SHAP bar plot for group and indiviudal first, and then claim how our overlapping graphs are helping?? Let me know your thoughts...}

\subsubsection{Visual Analysis of Overlapping SHAP Plots}

As we have described before, in the overlapping SHAP plots between the instance's SHAP values and the respective cases with the relative number of top features, we expect higher number of overlapping bars with the actual group. Here, we highlight five example cases from the three datasets- one FP case for $D_1$, one FN and one TN case for $D_2$, and one FN and one TP case for $D_3$. These queries address the \textbf{RQ1} and \textbf{RQ2}.

\noindent \textbf{Use Case Scenario Example from $D_1$ Dataset:} For a random positive prediction instance case in dataset $D_1$, Figure \ref{fig: Dataset_1_SHAP} presents the mapping of the individual prediction outcome into both FP and TP group SHAP bar plots and generated two overlapping bar graphs. It is evident from Figure \ref{fig: Dataset_1_SHAP}((a) and (b)) that this individual instance has higher number of overlapping bars with the FP group and very low number of overlapping bars with the TP group. Moreover, the raw prediction probability is $0.54$, which indicates the instance is a false-positive outcome and should be a treated as benign when taking any action by the analyst.  %his indicates that, the instance's SHAP values are quite similar with the FP mean SHAP values thus identify the prediction as an FP. Also, the predicted probability value is 0.540 which is way less than our considered threshold of 0.9. 

\begin{figure*}[!t]
\centering
\subcaptionbox{higher overlapping bars with FN\label{fig: fn_hikari_fn}}{\includegraphics[width=0.49\textwidth,height=6cm]{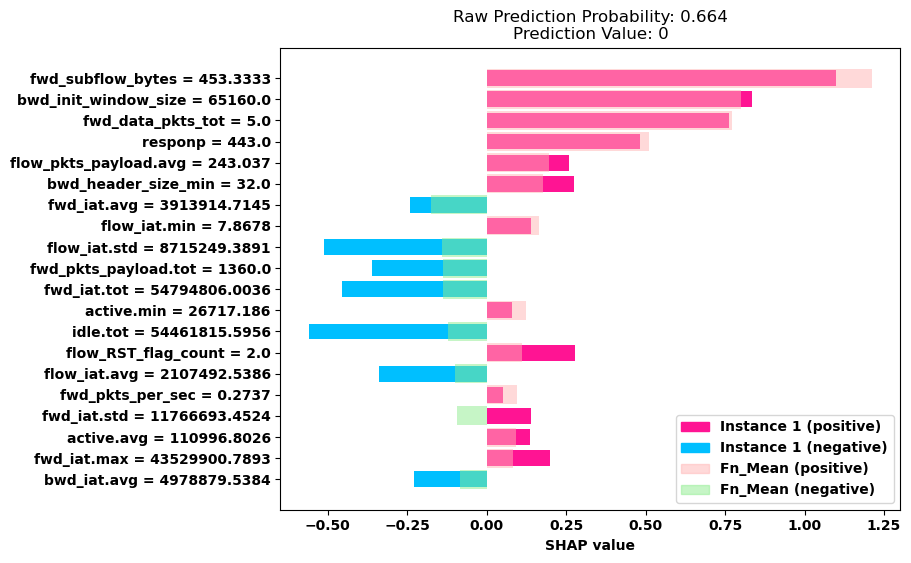}}%
\hfill % <-- Separation
\subcaptionbox{lower overlapping bars with TN\label{fig: fn_hikari_tn}}{\includegraphics[width=0.49\textwidth,height=6cm]{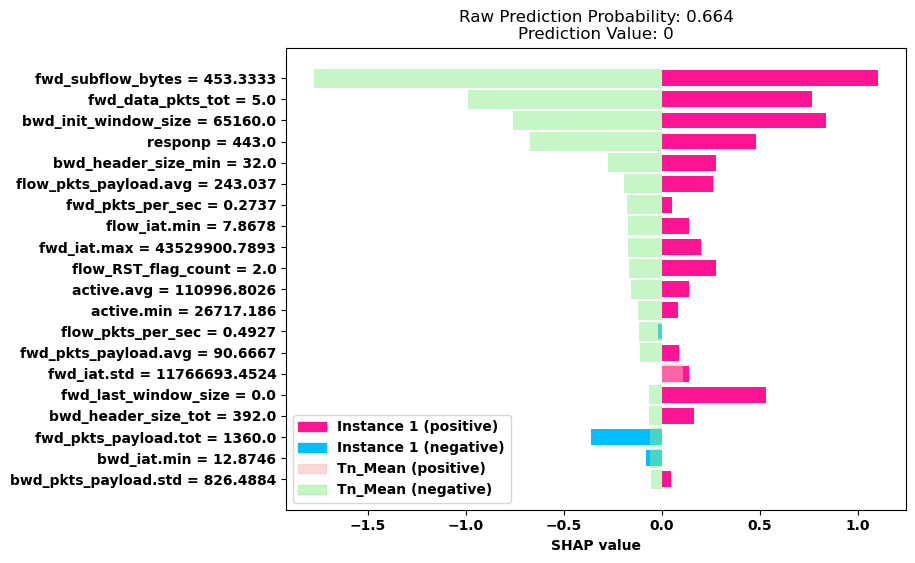}}%
\caption{Overlapping bar plots with a random false-negative instance within dataset $D_3$}
\label{fig: Dataset_3_SHAP}
\end{figure*}

\noindent \textbf{Use Case Scenario Example from $D_2$ Dataset:} 
For $D_2$, we have provided an example of a random negative ({\em benign}) prediction instance. We can see from Figure \ref{fig: Dataset_2_SHAP}((a) and (b)) that for a negative prediction, we map the individual features' SHAP values with the FN and TN group features where clearly the FN group has much more overlapping in this scenario compared to the TN group. Also, the raw prediction probability is $0.69$ and all these factors clearly indicating a false-negative prediction in this case. Thus, while taking any action by the analyst, they should identify this instance as false-negative and consider it as a real attack instance. We also provide an expected TN scenario (Figure \ref{fig: Dataset_2_SHAP_2}) where we still see plot similarity with the TN group rather with FN group with high probability value as well.  

\begin{figure*}[!t]
\centering
\subcaptionbox{equal number of overlapping bars with FP\label{fig: tp_hikari_fp}}{\includegraphics[width=0.50\textwidth,height=6cm]{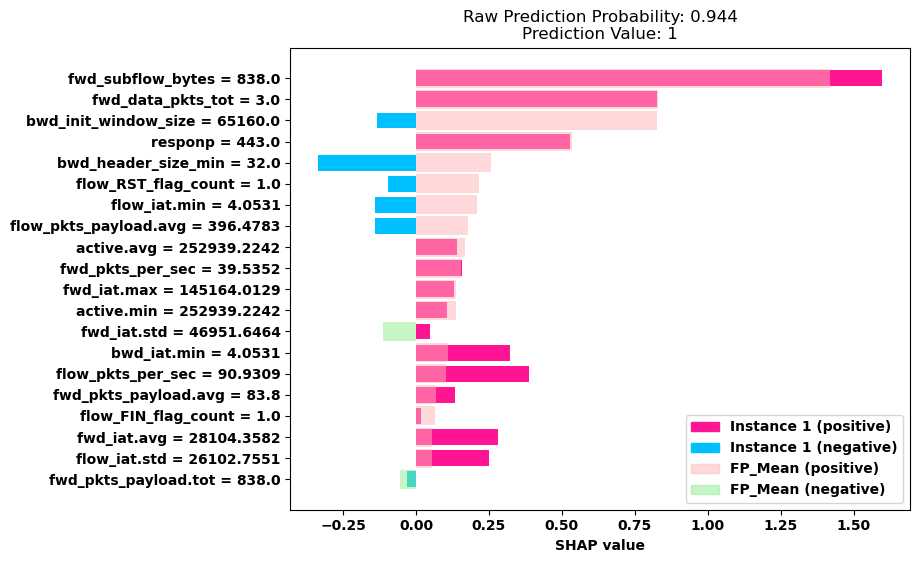}}%
\hfill % <-- Separation
\subcaptionbox{equal number of overlapping bars with TP\label{fig: tp_hikari_tp}}{\includegraphics[width=0.499\textwidth,height=6cm]{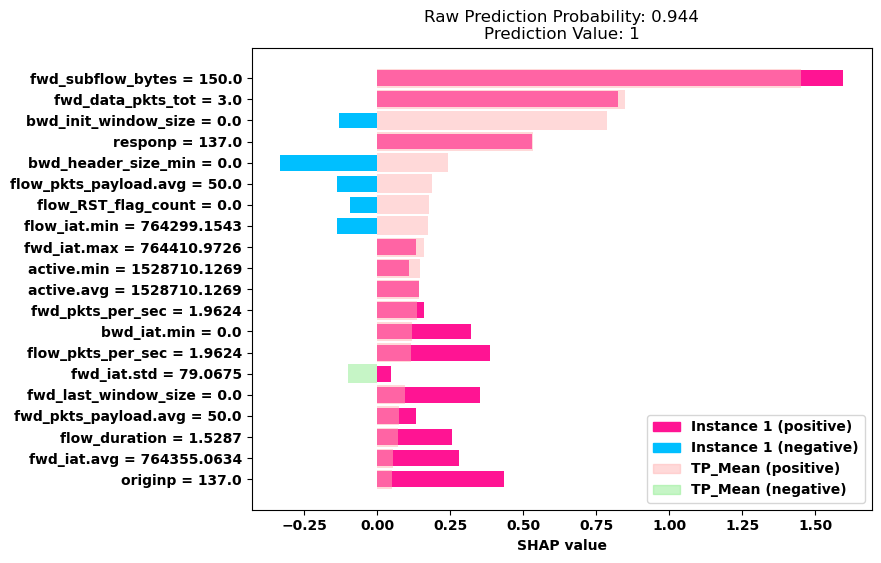}}%
\caption{Overlapping bar plots with a random true-positive instance within dataset $D_3$ (a case of confusion)}
\label{fig: Dataset_3_SHAP_2}
\end{figure*}

% case the  one with the FN mean SHAP value (figures \ref{fig: fn_nf_fn} and \ref{fig: fn_hikari_fn}) have the higher number of overlapping bars and another one with TN mean SHAP value (figures \ref{fig: fn_nf_tn} and \ref{fig: fn_hikari_tn}) has the lower number of overlapping bars. This indicates that, the instance's SHAP values are quite similar with the FN mean SHAP values thus this indicates the prediction as an FN. Also, the predicted probability values are 0.693 in figure \ref{fig: Dataset_2_SHAP} and 0.664 in figure \ref{fig: Dataset_3_SHAP} which are also less than our considered threshold of 0.9.

\noindent \textbf{Use Case Scenario Example from $D_3$ Dataset:} 
Again for the third dataset, $D_3$, we have provided an example of a random {\em benign} prediction instance. We have observed from Figure \ref{fig: Dataset_3_SHAP}((a) and (b)) that for a negative prediction, we map the individual features' SHAP values with the FN and TN group features where FN group has higher overlapping in this scenario compared to the TN group. In fact, most of the TN group feature contributions are in opposite direction (i.e., positively contributing top features are showing negative contribution for this instance). Also, the raw prediction probability is $0.664$. All these signs clearly indicating a false-negative prediction in this case and suggest the analyst to identify this instance as a real attack while taking any decision action.

\subsubsection{Evaluation of FP and FN Corrections}
We have further evaluated our proposed visual analysis approach with the end user experience where one computer science graduate student has played the role of an analyst without knowing the ground-truth. We provide the analyst with $300$ random instances ($100$ instances for each datasets) along with their overlapping SHAP bar graphs. The ultimate goal is to test how many of the instances can be correctly identified as {\em attack} versus {\em benign} and thus improve the overall model performance. In another way, we can say if something is false-positive and the analyst can correctly identify that, then the same action alike a truly {\em benign} sample can be taken for the false-positive instances. %The prediction probability threshold is set to be 0.9 as suggested from table  to ensure minimum possible loss in correct prediction and the ground truth is only known to the research group for further cross-analysis of the analyst result. 
In this case study scenario, we set the raw prediction probability threshold to $0.9$ to trust the model's outcome in case of confusion for taking a decision-- such an example is presented in Figure \ref{fig: Dataset_3_SHAP_2} where the overlapping plots are not decisive and may not give any clear indication. However, the raw probability of $0.94$ for {\em attack} prediction, in this case would recommend the analyst to regard the instance as an actual {\em attack}. This systematic approach for FP-FN identification along with decision-making criteria answers \textbf{RQ3}. %In case of any uncertainty in taking a decision where the raw-prediction probability we use the raw prediction probability threshold of $0.9$ to trust  (i.e. a use case scenario when raw prediction is below threshold but two plots are identical or any kind of hindrance) the analyst is suggested to acknowledge the model's prediction. 

\begin{table}[!ht]
    \centering
    \caption{Evaluation of Random Instances By Analyst}
    \vspace{-0.3em}
    \resizebox{0.45\textwidth}{!}
    % {\begin{tabular}{p{0.7cm}|p{0.35cm}p{0.6cm}|p{0.35cm}p{0.6cm}|p{0.35cm}p{0.6cm}|p{0.35cm}p{0.6cm}}
    {\begin{tabular}{c|cc|cc|cc|cc}
        \hline
        % First row (header)
        \multirow{2}{*}{\textbf{Dataset}} & \multicolumn{2}{c|}{\textbf{TP}} & \multicolumn{2}{c|}{\textbf{TN}} & \multicolumn{2}{c|}{\textbf{FP}} & \multicolumn{2}{c}{\textbf{FN}} \\
        \cline{2-9}
        % Second row (sub-headers)
        & \textbf{tested} & \textbf{correct} & \textbf{tested} & \textbf{correct} & \textbf{tested} & \textbf{correct} & \textbf{tested} & \textbf{correct} \\
        \hline
        % Sample data rows
        \textbf{$D_1$} & 35 & 35 & 35 & 34 & 15 & 13 & 15 & 5 \\
        \hline
        \textbf{$D_2$} & 35 & 33 & 35 & 35 & 15 & 11 & 15 & 9 \\
        \hline
        \textbf{ $D_3$} & 35 & 35 & 35 & 35 & 18 & 6 & 12 & 12 \\
        \hline
    \end{tabular}}
    \label{tab:analyst}
    \vspace{-1.2mm}
\end{table}

\vspace{-0.50mm}
Now, Table \ref{tab:analyst} shows the evaluation results for all three datasets where we clearly see a good number of correct identification of false-positive and false-negative instances to reduce FPR and FNR with very minimal impact on the TP and TN instances. Particularly, in dataset $D_1$ and $D_2$, the analyst has identified a very high percentage of FP instances. However, the analyst has struggled to identify FPs in $D_3$ (only 6 out 18) where all of the FN instances are correctly identified. Although the ratio of correct FP and FN detection is deviating in different dataset scenarios, we are able to detect a good amount of FP and FN instances across datasets showing the efficacy of our approach and this answers the \textbf{RQ4}.

%% file: discuss.tex
\vspace{-1mm}
This paper highlights how XAI based explanation graphs can enable more trust in IDS scenarios and aid analyst to make more reliable decision-making, specially against FP and FN instances. Even though we discuss our case study with binary classification, in case of multi-class classification the method can be applied as we compare the individual local explanation with the true positive group explanation for the certain class. The study still has the following limitations. %For example, if traffic instance ${\tt tf}_k$ is predicted as {\em `DDoS'}, then we compare and analyze the local explanation for ${\tt tf}_k$ with that of TP group for {\em DDoS} to identify if it is a false-positive identification or not. On the other hand, we assume true-negative as the truly predicted \textit{`benign'} cases in multi-class classification and false-negatives are those cases where any attack category traffic are classified as {\em benign}. For example, if traffic instance ${\tt tf}_k$ is predicted as \textit{`benign'}, then we compare and analyze the local explanation for ${\tt tf}_k$ with that of TN group to identify is it's a false-negative or not. However, this work still has few limitations discussed below. 

\vspace{-0.25mm}
\noindent \textbf{Limitations:} First, we have not considered deep learning models in this study %that are often used in literature with higher accuracy 
  as our primary focus have been more on the model explainability and transparency for decision-making. Second, the FPs and FNs mitigation process still needs human-analyst intervention through visual analysis of SHAP plots, which may introduces challenges like human error and well understanding of SHAP plots. %lack of domain-knowledge. limitations for lack of domain-knowledge. 
  Third, we only present case study with binary classification scenario, but multi-class classification may create complex plots scenarios.  %may  the study can be extended for multi-class classification scenarios. %In the future, it can be addressed by intelligent decision recommendation systems to aid analysts. 
  %\textbf{third}, even though the case study is presented using the IDS based datasets, we can generalize and replicate the methodological process for other dataset scenarios as well, which may have different explanation sets for FPs and TPs due to the change of dataset distribution but the idea of differences from FPs from the TPs or the FNs from the TNs will still hold. 
  Fourth, we have not considered adversarial attacks that can manipulate the XAI outputs and thus can manipulate the group-wise explanation for TPs and TNs, which can be explored in future studies. Fifth, if the model is not retrained frequently over time, then the current setup may not be effective in detecting FPs and FNs due to concept drift situations, which is not considered in the scope of this paper. %new incoming network traffic features that have different value ranges for top features. Hence, an online learning model may become a more viable and effective option in the future study.

%% file: conclusion.tex
\vspace{-1.5mm}
%may generate an invalid explanation scheme.
%\footnote{address the issue of using only 1 dataset, does not invalidate the process as the method is dataset independent}

%\footnote{address the adversarial attacks are not considered that can manipulate XAI, and can be studied in the future...}
%\footnote{how the method is still weak for new type of datapoint}
%\vspace{-0.8em}
%\subsection{Conclusion}
%\vspace{-0.45em}
In this paper, we propose an XAI-enabled approach with {\em XGBoost} model to accurately identify various intrusions or attack scenarios while also helping the analyst to correctly identify false positive and false negative through a more effective visual analysis approach. Our approach maps the local SHAP based feature explanation bar plots with the TP, FP, TN, FN group-wise explanation bar plots to generate a overlapping bar plot and find potential similarities and dissimilarities to correctly identify false positive and false negative instances. %We also showed that using a well-known state-of-the-art XAI method, {\em SHAP}, we can reduce the traditional false positive (FPs) and false negative (FNs) detection problems of any ML-based IDS. 
%We imperically showed that our methodology works well with binary (2-class), group-wise (8-class), and individual attack-wsie (34-class) classificaiton scenarios. 
Moreover, our case study with three independent datasets and their extensive evaluation presents the efficacy of our approach for reliable decision-making when dealing with false positive and false negative instances. %Overall, we feel similar methodological studies can be carried out with other relevant datasets to showcase the effectiveness of our methods in case of helping analyst in effective decision-making for intrusion detection. 
For reproducing the experiments, the detail implementation of the work can be found in the following public \href{https://anonymous.4open.science/r/Correcting-False-Positives-and-False-Negatives-in-IoT-Intrusion-Detection-Systems-with-XAI-2223/}{Github repository}.

%{\underline{here}} for reproducing the experiment results if needed.%\footnote{add this github page link here with anonymous account}  %we have implemented the Brier score concepts which gave us mathematical measurements of how much deviation exists between the predicted results and the actual result found in the test data portion. Finally, we have analyzed the feature contribution using SHAP and provided valuable insights on certain feature patterns and contributions to the model. Using this powerful tool, we also provide a direction on how to mitigate false positives and false negative predictions.

%In the future, we plan to incorporate robust deep learning-based models to increase the accuracy and using Deep Shap \cite{yuan2022empirical} to minimize the FPR and FNR at the earlier stage. Additionally, we also want to explore the automation engineering problem of identifying the FPs and FNs by creating an intelligent decision system based on the explanation strengths and weaknesses rather than depending on human-analyst driven data analysis.  %would be to automate the FP and FN prediction for a particular model with the help of XAI tools so that there is no need for immediate intervention of human users in further analyzing the result of the XAI tools.

%% file: appendix.tex
\begin{algorithm}[!h]
\small
\caption{SHAP-based Network Traffic Classification}
\label{alg:shap_algo}

\algsetup{linenosize=\small}
\begin{algorithmic}[1]
    \REQUIRE Trained ML model $M_m$ on $D_{train}$, $D_{test}$
    \STATE $D_{tp}, D_{tn}, D_{fp}, D_{fn} \gets M_m(D_{test})$

    \STATE $O_e \gets SHAP.explainer(M_m, D_{train})$
    \STATE $E \gets O_e( D_{test})$ \COMMENT{Global Mean SHAP Assignment}

    \STATE $E_{tp\_mean} \gets  E[D_{tp}]$
    \STATE $E_{tn\_mean} \gets  E[D_{tn}]$
    \STATE $E_{fp\_mean} \gets  E[D_{fp}]$ 
    \STATE $E_{fn\_mean} \gets  E[D_{fn}]$

    \FOR {\textbf{each} new incoming traffic instance $tf_i$}

        \STATE $Y_{i} \gets predict(M_m, tf_i)$, $P_i \gets predict\_prob(M_m,tf_i)$
    
        \STATE $E_{i} \gets O_e( tf_i)$ \COMMENT {$i$-th instance SHAP value assignment}

        \IF{$Y_i = attack$}
        \STATE Generate-Plot($E_i$, $E_{tp\_mean}$)
        \STATE note $plot\_sim_{tp}$
        \STATE Generate-Plot($E_i$, $E_{fp\_mean}$)
        \STATE note $plot\_sim_{fp}$
        %\COMMENT{Generate  two overlapping bar plots} %of $E_i$ with $E_{tp\_mean}$ and $E_{fp\_mean}$

        \IF{$P_i \geq threshold$ \OR  $plot\_sim_{tp} \geq plot\_sim_{fp}$}
        \STATE yield to $Y_i$

        \ELSE
        \STATE FP case detected
        
        \ENDIF

        \ELSE
        \STATE Generate-Plot($E_i$, $E_{tn\_mean}$)
        \STATE note $plot\_sim_{tn}$
        \STATE Generate-Plot($E_i$, $E_{fn\_mean}$)
        \STATE note $plot\_sim_{fn}$
        %\COMMENT{Generate  two overlapping bar plots} %of $E_i$ with $E_{tp\_mean}$ and $E_{fp\_mean}$
        \IF{$P_i \geq threshold$ \OR $plot\_sim_{tn} \geq plot\_sim_{fn}$}
        \STATE yield to $Y_i$

        \ELSE
        \STATE FN case detected
        
        \ENDIF
        %\STATE Generate  two overlapping bar plots of $E_i$ with $E_{tn\_mean}$ and $E_{fn\_mean}$

        \ENDIF
        
        % \IF{$P_i >$ threshold}
        % \IF{}
        %     \STATE Choose the case with the higher number of overlapping bars
        % \ELSE
        %     \STATE choose original $Y_i$
        % \ENDIF
    \ENDFOR
\end{algorithmic}
\end{algorithm}

%% file: main.bbl
\begin{thebibliography}{10}
\providecommand{\url}[1]{#1}
\csname url@samestyle\endcsname
\providecommand{\newblock}{\relax}
\providecommand{\bibinfo}[2]{#2}
\providecommand{\BIBentrySTDinterwordspacing}{\spaceskip=0pt\relax}
\providecommand{\BIBentryALTinterwordstretchfactor}{4}
\providecommand{\BIBentryALTinterwordspacing}{\spaceskip=\fontdimen2\font plus
\BIBentryALTinterwordstretchfactor\fontdimen3\font minus \fontdimen4\font\relax}
\providecommand{\BIBforeignlanguage}[2]{{%
\expandafter\ifx\csname l@#1\endcsname\relax
\typeout{** WARNING: IEEEtran.bst: No hyphenation pattern has been}%
\typeout{** loaded for the language `#1'. Using the pattern for}%
\typeout{** the default language instead.}%
\else
\language=\csname l@#1\endcsname
\fi
#2}}
\providecommand{\BIBdecl}{\relax}
\BIBdecl

\bibitem{IDS1}
H.-J. Liao, C.-H. {Richard Lin}, Y.-C. Lin, and K.-Y. Tung, ``Intrusion detection system: A comprehensive review,'' \emph{Journal of Network and Computer Applications}, vol.~36, no.~1, pp. 16--24, 2013.

\bibitem{IDS-attacks}
T.~M. Booij, I.~Chiscop, E.~Meeuwissen, N.~Moustafa, and F.~T. H.~d. Hartog, ``Ton\_iot: The role of heterogeneity and the need for standardization of features and attack types in iot network intrusion data sets,'' \emph{IEEE Internet of Things Journal}, vol.~9, no.~1, pp. 485--496, 2022.

\bibitem{Ontologybased}
F.~E. Ayo, J.~B. Awotunde, L.~A. Ogundele, O.~O. Solanke, B.~Brahma, R.~Panigrahi, and A.~K. Bhoi, ``Ontology-based layered rule-based network intrusion detection system for cybercrimes detection,'' \emph{Knowledge and Information Systems}, Feb 2024.

\bibitem{mir_host_loganalysis_2017}
M.~M.~A. Pritom, C.~Li, B.~Chu, and X.~Niu, ``A study on log analysis approaches using sandia dataset,'' in \emph{2017 26th International Conference on Computer Communication and Networks (ICCCN)}, 2017, pp. 1--6.

\bibitem{XAI_IDS_7}
\BIBentryALTinterwordspacing
E.~Lee, Y.~Lee, and T.~Lee, ``Automatic false alarm detection based on xai and reliability analysis,'' \emph{Applied Sciences}, vol.~12, no.~13, 2022. [Online]. Available: \url{https://www.mdpi.com/2076-3417/12/13/6761}
\BIBentrySTDinterwordspacing

\bibitem{XAI_IDS_8}
K.~Fujita, T.~Shibahara, D.~Chiba, M.~Akiyama, and M.~Uchida, ``Objection!: Identifying misclassified malicious activities with xai,'' in \emph{ICC 2022 - IEEE International Conference on Communications}, 2022, pp. 2065--2070.

\bibitem{pritom2022supporting}
M.~M.~A. Pritom and S.~Xu, ``Supporting law-enforcement to cope with blacklisted websites: Framework and case study,'' in \emph{2022 IEEE Conference on Communications and Network Security (CNS)}.\hskip 1em plus 0.5em minus 0.4em\relax IEEE, 2022, pp. 181--189.

\bibitem{shap}
S.~M. Lundberg, G.~Erion, H.~Chen, A.~DeGrave, J.~M. Prutkin, B.~Nair, R.~Katz, J.~Himmelfarb, N.~Bansal, and S.-I. Lee, ``From local explanations to global understanding with explainable ai for trees,'' \emph{Nature Machine Intelligence}, vol.~2, no.~1, pp. 2522--5839, 2020.

\bibitem{rufibach2010use_brier}
K.~Rufibach, ``Use of brier score to assess binary predictions,'' \emph{Journal of clinical epidemiology}, vol.~63, no.~8, pp. 938--939, 2010.

\bibitem{nimbalkar2021analysis}
P.~Nimbalkar and D.~Kshirsagar, ``Analysis of rule-based classifiers for ids in iot,'' in \emph{Data Science and Security: Proceedings of IDSCS 2021}.\hskip 1em plus 0.5em minus 0.4em\relax Springer, 2021, pp. 461--467.

\bibitem{ferrag2020rdtids}
M.~A. Ferrag, L.~Maglaras, A.~Ahmim, M.~Derdour, and H.~Janicke, ``Rdtids: Rules and decision tree-based intrusion detection system for internet-of-things networks,'' \emph{Future internet}, vol.~12, no.~3, p.~44, 2020.

\bibitem{kumar2021uids}
V.~Kumar, A.~K. Das, and D.~Sinha, ``Uids: a unified intrusion detection system for iot environment,'' \emph{Evolutionary intelligence}, vol.~14, no.~1, pp. 47--59, 2021.

\bibitem{ml_dl_ids}
Z.~Ahmad, A.~Shahid~Khan, C.~Wai~Shiang, J.~Abdullah, and F.~Ahmad, ``Network intrusion detection system: A systematic study of machine learning and deep learning approaches,'' \emph{Transactions on Emerging Telecommunications Technologies}, vol.~32, no.~1, p. e4150, 2021.

\bibitem{islam2021towards}
N.~Islam, F.~Farhin, I.~Sultana, M.~S. Kaiser, M.~S. Rahman, M.~Mahmud, A.~SanwarHosen, and G.~H. Cho, ``Towards machine learning based intrusion detection in iot networks,'' \emph{Computers, Materials \& Continua}, vol.~69, no.~2, 2021.

\bibitem{saheed2022machine}
Y.~K. Saheed, A.~I. Abiodun, S.~Misra, M.~K. Holone, and R.~Colomo-Palacios, ``A machine learning-based intrusion detection for detecting internet of things network attacks,'' \emph{Alexandria Engineering Journal}, vol.~61, no.~12, pp. 9395--9409, 2022.

\bibitem{verma2020machine}
A.~Verma and V.~Ranga, ``Machine learning based intrusion detection systems for iot applications,'' \emph{Wireless Personal Communications}, vol. 111, no.~4, pp. 2287--2310, 2020.

\bibitem{Khan2024}
M.~M. Khan and M.~Alkhathami, ``Anomaly detection in iot-based healthcare: machine learning for enhanced security,'' \emph{Scientific Reports}, vol.~14, no.~1, p. 5872, Mar 2024.

\bibitem{ge2019deep}
M.~Ge, X.~Fu, N.~Syed, Z.~Baig, G.~Teo, and A.~Robles-Kelly, ``Deep learning-based intrusion detection for iot networks,'' in \emph{2019 IEEE 24th pacific rim international symposium on dependable computing (PRDC)}.\hskip 1em plus 0.5em minus 0.4em\relax IEEE, 2019, pp. 256--25\,609.

\bibitem{khan2021deep}
M.~A. Khan, M.~A. Khan, S.~U. Jan, J.~Ahmad, S.~S. Jamal, A.~A. Shah, N.~Pitropakis, and W.~J. Buchanan, ``A deep learning-based intrusion detection system for mqtt enabled iot,'' \emph{Sensors}, vol.~21, no.~21, p. 7016, 2021.

\bibitem{electronics13061053_2024}
\BIBentryALTinterwordspacing
S.~Yaras and M.~Dener, ``Iot-based intrusion detection system using new hybrid deep learning algorithm,'' \emph{Electronics}, vol.~13, no.~6, 2024. [Online]. Available: \url{https://www.mdpi.com/2079-9292/13/6/1053}
\BIBentrySTDinterwordspacing

\bibitem{Jony_Arnob_2024}
\BIBentryALTinterwordspacing
A.~I. Jony and A.~K.~B. Arnob, ``A long short-term memory based approach for detecting cyber attacks in iot using cic-iot2023 dataset,'' \emph{Journal of Edge Computing}, Jan. 2024. [Online]. Available: \url{https://acnsci.org/journal/index.php/jec/article/view/648}
\BIBentrySTDinterwordspacing

\bibitem{hybrid_ids}
S.~Smys, A.~Basar, H.~Wang \emph{et~al.}, ``Hybrid intrusion detection system for internet of things (iot),'' \emph{Journal of ISMAC}, vol.~2, no.~04, pp. 190--199, 2020.

\bibitem{sadikin2020zigbee}
F.~Sadikin and S.~Kumar, ``Zigbee iot intrusion detection system: A hybrid approach with rule-based and machine learning anomaly detection.'' in \emph{IoTBDS}, 2020, pp. 57--68.

\bibitem{saif2022hiids}
S.~Saif, P.~Das, S.~Biswas, M.~Khari, and V.~Shanmuganathan, ``Hiids: Hybrid intelligent intrusion detection system empowered with machine learning and metaheuristic algorithms for application in iot based healthcare,'' \emph{Microprocessors and Microsystems}, p. 104622, 2022.

\bibitem{CIC-IDS-2017}
\BIBentryALTinterwordspacing
I.~Sharafaldin, A.~H. Lashkari, and A.~A. Ghorbani, ``Toward generating a new intrusion detection dataset and intrusion traffic characterization,'' in \emph{International Conference on Information Systems Security and Privacy}, 2018. [Online]. Available: \url{https://api.semanticscholar.org/CorpusID:4707749}
\BIBentrySTDinterwordspacing

\bibitem{CIC_18}
J.~L. Leevy and T.~M. Khoshgoftaar, ``A survey and analysis of intrusion detection models based on cse-cic-ids2018 big data,'' \emph{Journal of Big Data}, vol.~7, pp. 1--19, 2020.

\bibitem{CIC22}
S.~Dadkhah, H.~Mahdikhani, P.~K. Danso, A.~Zohourian, K.~A. Truong, and A.~A. Ghorbani, ``Towards the development of a realistic multidimensional iot profiling dataset,'' in \emph{2022 19th Annual International Conference on Privacy, Security \& Trust (PST)}, 2022, pp. 1--11.

\bibitem{CICIoT-2023-datasetpaper}
\BIBentryALTinterwordspacing
E.~C.~P. Neto, S.~Dadkhah, R.~Ferreira, A.~Zohourian, R.~Lu, and A.~A. Ghorbani, ``Ciciot2023: A real-time dataset and benchmark for large-scale attacks in iot environment,'' \emph{Sensors}, vol.~23, no.~13, 2023. [Online]. Available: \url{https://www.mdpi.com/1424-8220/23/13/5941}
\BIBentrySTDinterwordspacing

\bibitem{gan_ids_paper}
J.~Lee and K.~Park, ``Gan-based imbalanced data intrusion detection system,'' \emph{Personal and Ubiquitous Computing}, vol.~25, no.~1, pp. 121--128, 2021.

\bibitem{XAI_Decision}
C.~Shand, R.~Fong, and U.~Butt, ``How explainable artificial intelligence (xai) models can be used within intrusion detection systems (ids) to enhance an analyst's trust and understanding,'' in \emph{International Conference on Global Security, Safety, and Sustainability}.\hskip 1em plus 0.5em minus 0.4em\relax Springer, 2023, pp. 321--342.

\bibitem{IDS_XAI_2}
B.~Sharma, L.~Sharma, C.~Lal, and S.~Roy, ``Explainable artificial intelligence for intrusion detection in iot networks: A deep learning based approach,'' \emph{Expert Systems with Applications}, vol. 238, p. 121751, 2024.

\bibitem{IDS_XAI_3}
M.~Siganos, P.~Radoglou-Grammatikis, I.~Kotsiuba, E.~Markakis, I.~Moscholios, S.~Goudos, and P.~Sarigiannidis, ``Explainable ai-based intrusion detection in the internet of things,'' in \emph{Proceedings of the 18th International Conference on Availability, Reliability and Security}, ser. ARES '23.\hskip 1em plus 0.5em minus 0.4em\relax New York, NY, USA: Association for Computing Machinery, 2023.

\bibitem{XAI_IDS_mahmoud23}
M.~C. Gaitan-Cardenas, M.~Abdelsalam, and K.~Roy, ``Explainable ai-based intrusion detection systems for cloud and iot,'' in \emph{2023 32nd International Conference on Computer Communications and Networks (ICCCN)}, 2023, pp. 1--7.

\bibitem{kalakoti2024improving}
R.~Kalakoti, H.~Bahsi, and S.~N{\~o}mm, ``Improving iot security with explainable ai: Quantitative evaluation of explainability for iot botnet detection,'' \emph{IEEE Internet of Things Journal}, 2024.

\bibitem{exai_ieee_access_2024}
O.~Arreche, T.~R. Guntur, J.~W. Roberts, and M.~Abdallah, ``E-xai: Evaluating black-box explainable ai frameworks for network intrusion detection,'' \emph{IEEE Access}, vol.~12, pp. 23\,954--23\,988, 2024.

\bibitem{OmarAbdel_2023_XAI_IoT_Intrusion}
S.~Arisdakessian, O.~Wahab, A.~Mourad, H.~Otrok, and M.~Guizani, ``\BIBforeignlanguage{British English}{A survey on iot intrusion detection: Federated learning, game theory, social psychology, and explainable ai as future directions},'' \emph{\BIBforeignlanguage{British English}{IEEE Internet of Things Journal}}, vol.~10, no.~5, pp. 4059--4092, Mar. 2023.

\bibitem{2024IEEE_XAI-IoT}
A.~{Namrita Gummadi}, J.~C. {Napier}, and M.~{Abdallah}, ``{XAI-IoT: An Explainable AI Framework for Enhancing Anomaly Detection in IoT Systems},'' \emph{IEEE Access}, vol.~12, pp. 71\,024--71\,054, Jan. 2024.

\bibitem{Abououf2024ExplainableAF}
\BIBentryALTinterwordspacing
M.~Abououf, S.~Singh, R.~Mizouni, and H.~Otrok, ``Explainable ai for event and anomaly detection and classification in healthcare monitoring systems,'' \emph{IEEE Internet of Things Journal}, vol.~11, pp. 3446--3457, 2024. [Online]. Available: \url{https://api.semanticscholar.org/CorpusID:260046664}
\BIBentrySTDinterwordspacing

\bibitem{IDS_FP_reduce}
R.~da~Silveira~Lopes, J.~C. Duarte, and R.~R. Goldschmidt, ``False positive identification in intrusion detection using xai,'' \emph{IEEE Latin America Transactions}, vol.~21, no.~6, pp. 745--751, 2023.

\bibitem{Kim2021Cost}
H.~Kim, Y.~Lee, E.~Lee, and T.~Lee, ``Cost-effective valuable data detection based on the reliability of artificial intelligence,'' \emph{IEEE Access}, vol.~9, pp. 108\,959--108\,974, 2021.

\bibitem{XAI_HEALTH_FP_FN_SHAP}
O.~O. Bifarin, ``Interpretable machine learning with tree-based shapley additive explanations: Application to metabolomics datasets for binary classification,'' \emph{Plos one}, vol.~18, no.~5, p. e0284315, 2023.

\bibitem{xNIDS_2023_usenix}
\BIBentryALTinterwordspacing
F.~Wei, H.~Li, Z.~Zhao, and H.~Hu, ``{xNIDS}: Explaining deep learning-based network intrusion detection systems for active intrusion responses,'' in \emph{32nd USENIX Security Symposium (USENIX Security 23)}.\hskip 1em plus 0.5em minus 0.4em\relax Anaheim, CA: USENIX Association, Aug. 2023, pp. 4337--4354. [Online]. Available: \url{https://www.usenix.org/conference/usenixsecurity23/presentation/wei-feng}
\BIBentrySTDinterwordspacing

\bibitem{CADE_2021_usenix}
\BIBentryALTinterwordspacing
L.~Yang, W.~Guo, Q.~Hao, A.~Ciptadi, A.~Ahmadzadeh, X.~Xing, and G.~Wang, ``{CADE}: Detecting and explaining concept drift samples for security applications,'' in \emph{30th USENIX Security Symposium (USENIX Security 21)}.\hskip 1em plus 0.5em minus 0.4em\relax USENIX Association, Aug. 2021, pp. 2327--2344. [Online]. Available: \url{https://www.usenix.org/conference/usenixsecurity21/presentation/yang-limin}
\BIBentrySTDinterwordspacing

\bibitem{DeepAID}
\BIBentryALTinterwordspacing
D.~Han, Z.~Wang, W.~Chen, Y.~Zhong, S.~Wang, H.~Zhang, J.~Yang, X.~Shi, and X.~Yin, ``Deepaid: Interpreting and improving deep learning-based anomaly detection in security applications,'' in \emph{Proceedings of the 2021 ACM SIGSAC Conference on Computer and Communications Security}, ser. CCS ’21.\hskip 1em plus 0.5em minus 0.4em\relax ACM, Nov. 2021. [Online]. Available: \url{http://dx.doi.org/10.1145/3460120.3484589}
\BIBentrySTDinterwordspacing

\bibitem{Khanday_Fatima_Rakesh_2023}
S.~Khanday, H.~Fatima, and N.~Rakesh, ``A novel data preprocessing model for lightweight sensory iot intrusion detection,'' \emph{International Journal of Mathematical, Engineering and Management Sciences}, vol.~9, pp. 188--204, 02 2024.

\bibitem{linardatos2020explainable}
P.~Linardatos, V.~Papastefanopoulos, and S.~Kotsiantis, ``Explainable ai: A review of machine learning interpretability methods,'' \emph{Entropy}, vol.~23, no.~1, p.~18, 2020.

\bibitem{dt_paper}
J.~R. Quinlan, ``Induction of decision trees,'' \emph{Machine learning}, vol.~1, pp. 81--106, 1986.

\bibitem{xgb_paper}
\BIBentryALTinterwordspacing
T.~Chen and C.~Guestrin, ``Xgboost: A scalable tree boosting system,'' in \emph{Proceedings of the 22nd ACM SIGKDD International Conference on Knowledge Discovery and Data Mining}, ser. KDD ’16.\hskip 1em plus 0.5em minus 0.4em\relax ACM, Aug. 2016. [Online]. Available: \url{http://dx.doi.org/10.1145/2939672.2939785}
\BIBentrySTDinterwordspacing

\bibitem{rf_paper}
L.~Breiman, ``Random forests,'' \emph{Machine learning}, vol.~45, pp. 5--32, 2001.

\bibitem{sarhan2023nf}
M.~Sarhan, S.~Layeghy, and M.~Portmann, ``Nf-uq-nids-v2,'' 2023.

\bibitem{ferriyan2021generating}
A.~Ferriyan, A.~H. Thamrin, K.~Takeda, and J.~Murai, ``Generating network intrusion detection dataset based on real and encrypted synthetic attack traffic,'' \emph{applied sciences}, vol.~11, no.~17, p. 7868, 2021.

\bibitem{Gheni2024-as}
H.~Q. Gheni and W.~L. Al-Yaseen, ``Two-step data clustering for improved intrusion detection system using ciciot2023 dataset,'' Mar 2024.

\bibitem{vishwakarma2022dids}
M.~Vishwakarma and N.~Kesswani, ``Dids: A deep neural network based real-time intrusion detection system for iot,'' \emph{Decision Analytics Journal}, vol.~5, p. 100142, 2022.

\bibitem{sarhan2021netflow}
M.~Sarhan, S.~Layeghy, N.~Moustafa, and M.~Portmann, ``Netflow datasets for machine learning-based network intrusion detection systems,'' in \emph{Big Data Technologies and Applications: 10th EAI International Conference, BDTA 2020, and 13th EAI International Conference on Wireless Internet, WiCON 2020, Virtual Event, December 11, 2020, Proceedings 10}.\hskip 1em plus 0.5em minus 0.4em\relax Springer, 2021, pp. 117--135.

\bibitem{adli2023anomaly}
T.~B. Adli, S.-B.~B. Amokrane, B.~Z. Pavlovi{\'c}, M.~Z.~M. Laidouni, and T.-e. A.~A. Benyahia, ``Anomaly network intrusion detection system based on netflow using machine/deep learning,'' \emph{Vojnotehni{\v{c}}ki glasnik}, vol.~71, no.~4, pp. 941--969, 2023.

\bibitem{vitorino2024reliable}
J.~Vitorino, M.~Silva, E.~Maia, and I.~Pra{\c{c}}a, ``Reliable feature selection for adversarially robust cyber-attack detection,'' \emph{Annals of Telecommunications}, pp. 1--15, 2024.

\bibitem{fernandes2022network}
R.~Fernandes and N.~Lopes, ``Network intrusion detection packet classification with the hikari-2021 dataset: a study on ml algorithms,'' in \emph{2022 10th International Symposium on Digital Forensics and Security (ISDFS)}.\hskip 1em plus 0.5em minus 0.4em\relax IEEE, 2022, pp. 1--5.

\end{thebibliography}
